\documentclass[10pt]{IEEEtran}
\usepackage{epsfig}
\usepackage{amsmath,graphicx,subfigure}
\usepackage{tikz}
\usetikzlibrary{decorations.pathreplacing}
\usetikzlibrary{dsp,chains,arrows,shapes,spy}
\usepackage{caption}
\usepackage{enumitem}
\tikzset{
    font={\fontsize{9}{11.0476pt}\selectfont}}
\usepackage{mathtools,amssymb,cuted}
\usepackage{pgfplots}
\pgfplotsset{compat=newest}
\usepackage{romannum}
\usepackage{algpseudocode}
\usepackage{algorithm}
\algnewcommand{\Parameters}[1]{%
    \State \textbf{\underline{Parameters}:}
    \Statex \hspace*{\algorithmicindent}\parbox[t]{.8\linewidth}{\raggedright #1}
}

\algnewcommand{\Input}[1]{%
    \State \textbf{\underline{Inputs}:}
    \Statex \hspace*{\algorithmicindent}\parbox[t]{.8\linewidth}{\raggedright #1}
}

\algnewcommand{\Initialization}[1]{%
    \State \textbf{\underline{Initialization}:}
    \Statex \hspace*{\algorithmicindent}\parbox[t]{.8\linewidth}{\raggedright #1}
}

\newcommand\undermat[2]{%
    \makebox[0pt][l]{$\smash{\underbrace{\phantom{%
                    \begin{matrix}#2\end{matrix}}}_{\text{$#1$}}}$}#2}
\begin{document}

\title{Dynamic Oversampling Tecniques for 1-Bit ADCs in Large-Scale MIMO Systems}

\author{Zhichao~Shao,~\IEEEmembership{Student Member,~IEEE,}
    Lukas~T.~N.~Landau,~\IEEEmembership{Member,~IEEE,}
    and~Rodrigo~C.~de~Lamare,~\IEEEmembership{Senior~Member,~IEEE}
    \vspace{-2em}
    \thanks{The authors are with the Pontifical Catholic
    University of Rio de Janeiro, Centre for Telecommunications Studies,
    Rio de Janeiro, CEP 22453-900, Brazil (e-mail: {zhichao.shao;lukas.landau;delamare}@cetuc.puc-rio.br).
    This work has been supported by CNPq, FAPERJ, ELIOT ANR-18-CE40-0030 and FAPESP 2018/12579-7 and 2015/24499-0.}}

\maketitle

\begin{abstract}
In this work, we investigate dynamic oversampling techniques for
large-scale multiple-antenna systems equipped with low-cost and
low-power 1-bit analog-to-digital converters at the base stations.
To compensate for the performance loss caused by the coarse
quantization, oversampling is applied at the receiver. Unlike
existing works that use uniform oversampling, which samples the
signal at a constant rate, a novel dynamic oversampling scheme is
proposed. The basic idea is to perform time-varying nonuniform
oversampling, which selects samples with nonuniform patterns that
vary over time. We consider two system design criteria: a design
that maximizes the achievable sum rate and another design that
minimizes the mean square error of detected symbols. Dynamic
oversampling is carried out using a dimension reduction matrix
$\mathbf{\Delta}$, which can be computed by the generalized
eigenvalue decomposition or by novel submatrix-level feature
selection algorithms. Moreover, the proposed scheme is analyzed in
terms of convergence, computational complexity and power consumption
at the receiver. Simulations show that systems with the proposed
dynamic oversampling outperform those with uniform oversampling in
terms of computational cost, achievable sum rate and symbol error
rate performance.
\end{abstract}


\section{Introduction}
\IEEEPARstart{I}{n} the uplink of large-scale multiple-antenna or
MIMO (Multiple-Input Multiple-Output) systems, users are served by a
large number of antenna chains at the base stations (BS) over the
same time-frequency resource \cite{6736761}. Compared to standard
MIMO networks, large-scale MIMO can multiply the capacity of a
wireless connection without requiring more spectrum, which increases
the spectral efficiency \cite{6457363}. This substantial improvement
makes large-scale MIMO an essential component of fifth-generation
(5G) communication systems \cite{6736746,Andrews}. However, with a
large number of antennas, it may not be possible to deploy expensive
and power-hungry hardware at the BS (as in conventional
multiple-antenna BS). Each receive antenna at the BS is connected to
a radio-frequency (RF) chain, which mainly consists of two
analog-to-digital converters (ADCs), low noise amplifier (LNA),
mixers, automatic gain control (AGC) and filters. Among all these
components, the power consumption of ADCs dominates the total power
of the whole RF chain. The study in \cite{Walden} has shown that two
factors, namely, sampling rate and the number of quantization bits,
influence the power consumption of ADCs, where the latter is the
main factor. As illustrated in \cite{Walden}, the power consumption
of ADCs grows exponentially with the number of quantization bits.
The deployment of current commercial high-resolution (8-12 bits)
ADCs is unaffordable for the practical use of large-scale MIMO. To
alleviate this requirement, the use of ADCs with coarse quantization
(1-4 bits) can largely reduce the power consumption at the BS and is
more suitable for large-scale MIMO systems. In this work, an extreme
1-bit resolution case is considered, where the in-phase and
quadrature components of the received samples are separately
quantized to 1 bit. This solution is particularly attractive to
large-scale MIMO systems, since each of the RF chains only contains
simple analog comparators and there is no AGC \cite{5288814}. This
advantage can substantially decrease both the power consumption and
the hardware cost at the BS.

\subsection{Previous Works}

An inherent characteristic of the 1-bit ADCs is the non-linearity,
which often results in a large distortion of the input signal.
Numerous methods have been proposed in the literature to deal with
such non-linearity. The authors in
\cite{Jacobsson,7155570,mezghani2012capacity} have investigated the
sum rate and given capacity upper and lower bounds of quantized MIMO
systems. In addition to capacity characterization studies, channel
estimation \cite{Li,Stockle,8491063} and signal detection
\cite{7439790,Shao2,Jeon,8610159,8240630} have also been analyzed,
where the work in \cite{8610159} has been developed for
frequency-selective channels. In a massive MIMO downlink system with
1-bit digital-to-analog converters (DACs), the severe distortions
due to the coarse quantization can be partially recovered by using
appropriate precoding designs \cite{Saxena,Landau,8227757,8754755}.
However, most of the previously reported works operate over
frequency-flat fading channels, which are rarely encountered in
modern broadband wireless communication systems. Extensions to
frequency-selective channels were reported in
\cite{Studer,Mollen2,Zhang}.

Most works on 1-bit quantized systems operate at the
Nyquist-sampling rate, where only one sample is obtained in a
Nyquist interval. To increase the information rate, oversampling (or
faster-than-Nyquist signaling) can be applied so that more samples
are obtained in one Nyquist interval. The first work on 1-bit
quantized signals with oversampling has been reported in
\cite{265506}, which shows a great advantage in terms of the
achievable rate. For Gaussian noisy channels, the authors in
\cite{5662127} have demonstrated the advantage of oversampling from
a capacity viewpoint. Furthermore, in \cite{7022948} the authors
have investigated the influence of different pulse shaping filters
on the 1-bit quantized systems with oversampling. The results show
that when using root-raised-cosine (RRC) filters, information rates
can be increased.

Recently, several works have investigated 1-bit quantization with
uniform oversampling in MIMO systems. The study in \cite{8487039}
has shown that oversampling can provide a gain in signal-to-noise
ratio (SNR) of about 5dB for the same symbol error rate (SER) and
achievable rate with a linear zero forcing (ZF) receiver. Analytical
bounds on the SER and achievable rate were also derived. The works
in \cite{8683750,ce1b} have devised channel estimation algorithms
for such systems. To reduce the computational cost caused by the
extra samples resulting from oversampling, a sliding window
technique was proposed in \cite{8450809}, where each transmission
block is separated into several sub-blocks for further signal
processing.

\subsection{Contributions}

In this work, we propose dynamic oversampling techniques for
large-scale multiple-antenna systems with 1-bit ADCs at the receiver
whose preliminary results were reported in \cite{Shao_dynamic}. In
contrast to previous works that consider uniform oversampling
\cite{8487039}, a novel dynamic oversampling scheme is developed,
which consists of time-varying nonuniform oversampling strategies.
In the proposed dynamic oversampling scheme, two rates are
introduced, namely, an initial sampling rate and a signal processing
rate. The received signal is initially oversampled at a higher rate
and then processed by dimension reduction matrices, which either
combine or select samples prior to further signal processing. The
proposed dynamic oversampling is highly innovative because it
dynamically selects samples and extracts information from them in a
more effective way, resulting in significant performance gains over
existing uniform oversampling and Nyquist-rate sampling schemes. We
consider design criteria to combine or select samples based on the
maximization of the sum rate and the minimization of the mean square
error (MSE) of the detected symbols. For signal detection, we employ
the sliding window technique \cite{8450809} in the MSE-based design
along with a dynamic oversampling based low-resolution aware minimum
mean square error (LRA-MMSE) detector. Following oversampling at a
higher rate, dynamic oversampling is performed by a dimension
reduction matrix $\mathbf{\Delta}$ computed by the generalized
eigenvalue decomposition (GEVD) or by novel submatrix-level feature
selection (SL-FS) algorithms. The proposed SL-FS algorithms are
highly original as they perform dimension reduction using an
efficient sub-matrix level sample selection strategy that reduces
the computational cost by at least an order of magnitude, while
their performance obtained through simulations is satisfactory.
Moreover, we examine the proposed techniques in terms of
convergence, computational complexity and power consumption at the
receiver. The proposed dynamic oversampling and the existing uniform
oversampling and Nyquist sampling schemes are compared using the
proposed LRA-MMSE and existing receivers. Simulations show that
systems with the proposed dynamic oversampling outperform those with
uniform oversampling and Nyquist-rate sampling in terms of
computational cost, achievable sum rate and symbol error rate
performance.

The rest of this paper is organized as follows: in Section
\Romannum{2} the system model for 1-bit dynamic oversampled
large-scale MIMO is described. In Section \Romannum{3} we derive the
sum rate and MSE based system designs and illustrate the design
algorithms for obtaining the reduction matrix in Section
\Romannum{4}. In Section \Romannum{5}, the
convergence, computational cost and power consumption of the
proposed scheme are examined. Simulations are presented and
discussed in Section \Romannum{6} and the paper is concluded in
Section \Romannum{7}.

Notation: throughout the paper, bold letters indicate vectors and
matrices, non-bold letters express scalars. The operators
$(\cdot)^T$, $(\cdot)^H$, $E\{\cdot\}$, $(\cdot)^{-1}$ stand for the
transposition, Hermitian transposition, expectation and inverse,
respectively. $\mathbf{I}_n$ denotes $n\times n$ identity matrix and
$\mathbf{0}_n$ is a $n\times 1$ all zeros column vector.
Additionally, $\text{diag}(\mathbf{A})$ is a diagonal matrix only
containing the diagonal elements of $\mathbf{A}$ and
$\text{blkdiag}(\cdot)$  is a block matrix such that the
main-diagonal blocks are matrices and all off-diagonal blocks are
zero matrices. $\otimes$, $\text{det}(\cdot)$, $\text{Tr}(\cdot)$ is
denoted by the operation of Kronecker product, determinant and
trace, respectively. $\lfloor a\rfloor$ gets the largest integer
smaller or equal to a and $\mod(a,b)$ returns the remainder after
division of a by b. The notation and the main
variables used in this paper are summarized in Table
\ref{tab:notation}.

\begin{table}[!h]
    \centering
    \caption{Notation and main variables}
    \includegraphics{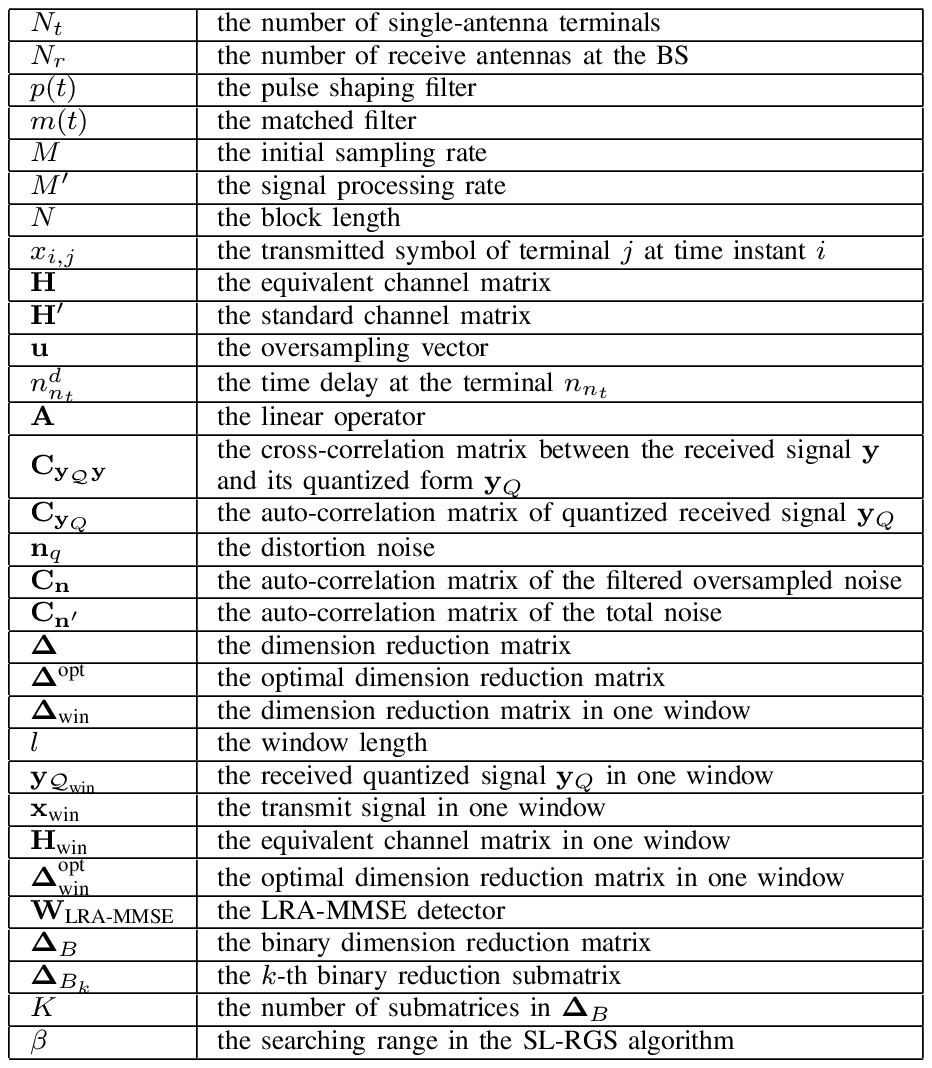}
    \label{tab:notation}
\end{table}

\section{System Model and Statistical Properties of 1-bit Quantization}

\begin{figure*}[!htb]
    \centering
    \includegraphics{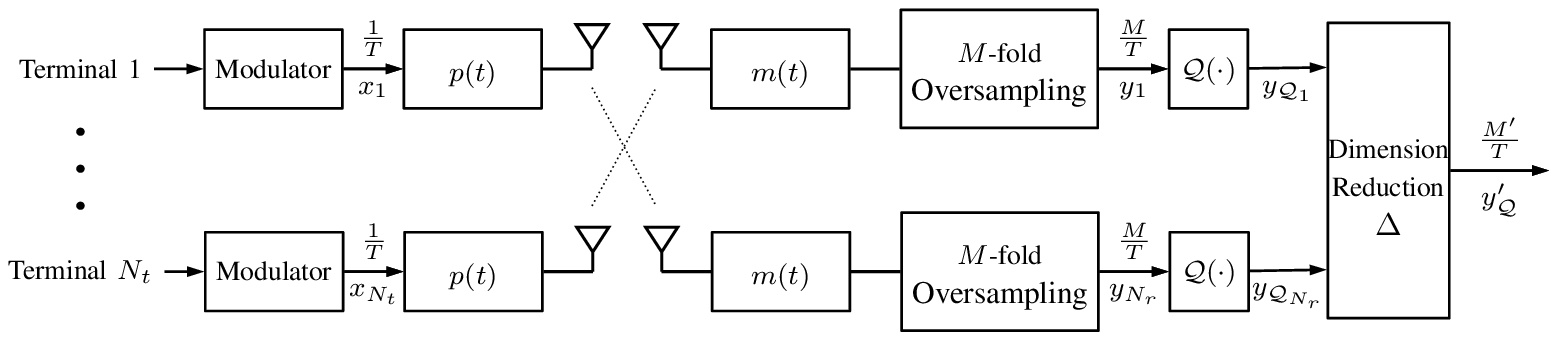}
    \caption{System model of multi-user multiple-antenna system with 1-bit ADCs and oversampling at the receiver.}
    \label{fig:transmitter}
\end{figure*}

The overall system model is illustrated with a block diagram in Fig.
\ref{fig:transmitter}, where the received oversampled signal
$\mathbf{y} \in \mathbb{C}^{MNN_r \times 1}$ for the single-cell
uplink large-scale multiple-antenna system with $N_t$ single-antenna
terminals and $N_r$ receive antennas ($N_r\gg N_t$) at the BS is
written as
\begin{equation}
\mathbf{y}=\mathbf{H}\mathbf{x}+\mathbf{n}.
\label{equ_model}
\end{equation}
In the block processing scheme, the initial sampling rate is a
factor $M$ times the Nyquist rate. After dimension reduction the
system is downsampled to a factor $M'$ times the Nyquist rate, named
signal processing rate, where $M\geq M'$. The vector $\mathbf{x}\in
\mathbb{C}^{NN_t \times 1}$ contains all transmitted symbols within
one block of $N$ data symbols, which is arranged as
\begin{equation}
\mathbf{x} = \left[x_{1,1} \quad \cdot\cdot\cdot \quad x_{N,1} \quad x_{1,2} \quad \cdot\cdot\cdot \quad x_{N,N_t}\right]^T,
\label{equ_x}
\end{equation}
where $x_{i,j}$ corresponds to the transmitted symbol of terminal
$j$ at time instant $i$ and is independently and identically
distributed (IID) with unit power $E[|x_{i,j}|^2]=1$.
Similar to (\ref{equ_x}), the received vector
after oversampling $\mathbf{y}$ has the form
\begin{equation}
    \mathbf{y} = \left[y_{1,1} \quad \cdot\cdot\cdot \quad y_{MN,1} \quad y_{1,2} \quad \cdot\cdot\cdot \quad y_{MN,N_r}\right]^T,
\end{equation}
where $y_{i,j}$ denotes the oversampled symbol of receiver
$j$ at time instant $i$. The vector $\mathbf{n}\in\mathbb{C}^{MNN_r \times 1}$ is the filtered
oversampled noise expressed by
\begin{equation}
\mathbf{n} = \left(\mathbf{I}_{N_r}\otimes \mathbf{G}\right)\mathbf{w},
\end{equation}
where $\mathbf{w}\sim
\mathcal{CN}\left(\mathbf{0}_{3MNN_r},\sigma^2_n\mathbf{I}_{3MNN_r}\right)$\footnote{Note that the noise samples are described such that each entry of
$\mathbf{n}$ has the same statistical properties. Since the receive
filter $m(t)$ has a length of $2MN + 1$ samples, $3MN$ unfiltered
noise samples in the noise vector $\mathbf{w}$ need to be considered
for the description of an interval of $MN$ samples of the filtered
noise $\mathbf{n}$.} contains IID complex Gaussian random variables
with zero mean and variance $\sigma^2_n$. The matrix
$\mathbf{G}\in\mathbb{R}^{MN\times 3MN}$ is a Toeplitz matrix
described by (\ref{equ:R}), where $m(t)$ represents the impulse response of
the matched filter at the time instant $t$.
\begin{figure*}[!htbp]
    \begin{equation}
    \mathbf{G} = \begin{bmatrix}
    m(-NT)& m(-NT+\frac{1}{M}T)& \dots& m(NT) & 0 & \dots & 0\\
    0 & m(-NT)& \dots & m(NT-\frac{1}{M}T) & m(NT) & \dots & 0\\
    \vdots & \vdots & \ddots & \vdots & \vdots & \ddots & \vdots\\
    0 & 0 & \dots & m(-NT)& m(-NT+\frac{1}{M}T)& \dots& m(NT)\\
    \end{bmatrix}
    \label{equ:R}
    \end{equation}
\end{figure*}
$T$ denotes one symbol period. We introduce the equivalent channel
matrix given by
\begin{equation}
\mathbf{H} = \left(\mathbf{H}'\otimes \mathbf{I}_{MN}\right)\text{blkdiag}\left([\mathbf{Z}_1,\dots,\mathbf{Z}_{N_t}]\right)(\mathbf{I}_{NN_t}\otimes\mathbf{u}),
\label{equ:H}
\end{equation}
where $\mathbf{H}'\in\mathbb{C}^{N_r \times N_t}$ is the standard
channel matrix prior to oversampling in which each coefficient
models the propagation effects between a user and an antenna
element. The vector $\mathbf{u}$ is employed as an oversampling
operator defined as the vector $\mathbf{u}$ with the size of $M
\times 1$
\begin{equation}
\mathbf{u} = \left[0 \quad \cdots \quad 0 \quad 1\right]^T.
\end{equation}
Similar to $\mathbf{G}$, the Toeplitz matrix
$\mathbf{Z}_{n_t}\in\mathbb{R}^{MN\times MN}$ contains the
impulse response of $z(t)$ at different time instants, where $z(t)$ is the
convolution of the transmit filter $p(t)$ and the receive filter
$m(t)$, and is given by
\begin{equation}
\resizebox{\columnwidth}{!}{$\displaystyle
    \mathbf{Z}_{n_t} = \begin{bmatrix}
    z[n^d_{n_t}] & z[n^d_{n_t}+\frac{T}{M}] & \dots & z[n^d_{n_t}+NT-\frac{1}{M}T]\\
    z[n^d_{n_t}-\frac{T}{M}] & z[n^d_{n_t}] & \dots & z[n^d_{n_t}+NT-\frac{2}{M}T]\\
    \vdots & \vdots & \ddots & \vdots\\
    z[n^d_{n_t}-NT+\frac{1}{M}T] & z[n^d_{n_t}-NT+\frac{2}{M}T] & \dots & z[n^d_{n_t}]\\
    \end{bmatrix}.$}
\end{equation}
In practical communication systems, the transmission delay is
unavoidable due to the different transmission paths of the users to
the BS. An asynchronous system is then considered by assuming the terminal
$n_t$ sends its signal to the BS delayed by the time $n^d_{n_t}$.
In this work, we adopt the Gaussian
assumption to simplify several signal processing tasks even though
they might not be accurate in practice. Moreover, the Bussgang
theorem \cite{Bussgang} has been extensively used to design and
analyze signal processing techniques that deal with coarsely
quantized signals thanks to the central limit theorem and the fact
that we are dealing with the superposition of multiple user signals.

Let $\mathcal{Q}(\cdot)$ represent the 1-bit quantization at the
receiver, the resulting quantized signal $\mathbf{y}_\mathcal{Q} \in
\mathbb{C}^{MNN_r \times 1}$ is given by
\begin{equation}
\mathbf{y}_\mathcal{Q}=\mathcal{Q}\left(\mathbf{y}\right)=\mathcal{Q}\left(\mathfrak{R}\{\mathbf{y}\}\right) + j\mathcal{Q}\left(\mathfrak{I}\{\mathbf{y}\}\right),
\label{system_model}
\end{equation}
where $\mathfrak{R}\{\cdot\}$ and $\mathfrak{I}\{\cdot\}$ get the
real and imaginary part, respectively. They are quantized to $\{\pm1\}$ and scaled to $\{\pm\frac{1}{\sqrt{2}}\}$ based on the sign.

The Bussgang theorem \cite{Bussgang} implies that
the output of the nonlinear quantizer can be decomposed into a
desired signal component and an uncorrelated distortion noise
$\mathbf{n}_q$:
\begin{equation}
\mathbf{y}_\mathcal{Q}=\mathbf{Ay}+\mathbf{n}_q,
\label{equ_linearBuss}
\end{equation}
where $\mathbf{A}\in\mathbb{R}^{MNN_r \times MNN_r}$ is the linear
operator chosen independently from $\mathbf{y}$, described by
\begin{equation}
\mathbf{A}=\mathbf{C}_{\mathbf{y}_{\mathcal{Q}}\mathbf{y}}\mathbf{C}_{\mathbf{y}}^{-1}
\label{equ_ap}
\end{equation}
and the distortion $\mathbf{n}_q$ has the covariance matrix
\begin{equation}
\mathbf{C}_{\mathbf{n}_q} = \mathbf{C}_{\mathbf{y}_{\mathcal{Q}}}-\mathbf{A}\mathbf{C}_{\mathbf{y}}\mathbf{A}^H.
\end{equation}
The matrix $\mathbf{C}_{\mathbf{y}_{\mathcal{Q}}\mathbf{y}}$ is the
cross-correlation matrix between the received signal $\mathbf{y}$
and its quantized form $\mathbf{y}_{\mathcal{Q}}$ that is given by
\begin{equation}
\mathbf{C}_{\mathbf{y}_{\mathcal{Q}}\mathbf{y}}=\sqrt{\frac{2}{\pi}}\mathbf{K}\mathbf{C}_\mathbf{y},\quad\text{with}\quad \mathbf{K} = \text{diag}(\mathbf{C}_{\mathbf{y}})^{-\frac{1}{2}}
\label{equ_Cyq}
\end{equation}
and the auto-correlation matrix
$\mathbf{C}_{\mathbf{y}_\mathcal{Q}}$ is obtained through the arcsin
law \cite{Jacovitti}, which results in
\begin{equation}
\mathbf{C}_{\mathbf{y}_\mathcal{Q}}=\frac{2}{\pi}\left(\text{sin}^{-1}(\mathbf{K}\mathbf{C}_{\mathbf{y}}^R\mathbf{K})+j\text{sin}^{-1}(\mathbf{K}\mathbf{C}_{\mathbf{y}}^I\mathbf{K})\right),
\label{equ_arcsin}
\end{equation}
where $\mathbf{C}_{\mathbf{y}}$ is calculated as
\begin{equation}
\begin{aligned}
\mathbf{C}_{\mathbf{y}}&=\mathbf{H}\mathbf{H}^H+\mathbf{C}_\mathbf{n}\\&=\mathbf{H}\mathbf{H}^H+\sigma_n^2(\mathbf{I}_{N_r}\otimes \mathbf{GG}^H).
\end{aligned}
\label{equ_Cy}
\end{equation}
Based on this decomposition, the signal after the 1-bit quantizer
can be written in the following form
\begin{equation}
\begin{aligned}
\mathbf{y}_\mathcal{Q}&=\mathbf{Ay}+\mathbf{n}_q\\&=\mathbf{AHx}+\mathbf{An}+\mathbf{n}_q.
\end{aligned}
\label{equ_linearBuss2}
\end{equation}

\section{System Design with Dynamic Oversampling}

Unlike the existing uniform oversampling, we devise an advanced
oversampling technique, named dynamic oversampling, where the system
is initially oversampled at a higher rate, a factor $M$ times the
Nyquist rate, and only a few samples are selected and further
processed. The proposed dynamic oversampling technique performs
time-varying nonuniform oversampling, which selects samples with
nonuniform patterns that vary over time. Uniform oversampling can be
interpreted as a special case of dynamic oversampling when $M=M'$
and the selection of samples is carried out using a time-invariant
uniform pattern. In this section, the proposed dynamic oversampling
technique is detailed based on two different design strategies: an
approach that maximizes the sum rate and another that minimizes the
MSE of detected symbols.

\subsection{Sum Rate based System Design}

Let us consider $\mathbf{n}' = \mathbf{An}+\mathbf{n}_q$ in
(\ref{equ_linearBuss2}) and assume it is Gaussian distributed with
the covariance matrix
\begin{equation}
    \mathbf{C}_{\mathbf{n}'}=\mathbf{A}\mathbf{C}_{\mathbf{n}}\mathbf{A}^H+\mathbf{C}_{\mathbf{n}_q}.
\end{equation}
The uplink sum rate lower bound\footnote{Note that the actual sum
rate is difficult to calculate due to the unknown characteristic of
the distortion noise $\mathbf{n}_q$. Similar to the work in
\cite{mezghani2012capacity}, $\mathbf{n}'$ is assumed to be Gaussian
distributed. This method minimizes the actual mutual information but
simplifies the analysis.} is then given by
\begin{equation}
\begin{aligned}
C_{LB}=E\left\{\frac{1}{N}\log_2\det\left(\mathbf{I}_{MNN_r}+\mathbf{AH}\mathbf{H}^H\mathbf{A}^H\mathbf{C}_{\mathbf{n}'}^{-1}\right)\right\}
\end{aligned},
\label{equ_cap}
\end{equation}
which can be simplified to
\begin{equation}
C_{LB}=E\left\{\frac{1}{N}\log_2\det\left(\mathbf{C}_{\mathbf{y}_{\mathcal{Q}}}\mathbf{C}_{\mathbf{n}'}^{-1}\right)\right\}.
\end{equation}
Assuming that the dimension reduction operation can be
mathematically described as a linear transformation with the matrix
$\mathbf{\Delta}$ \cite{4803749}, the received signal is then
reduced to
\begin{equation}
\mathbf{y}'_\mathcal{Q}=\mathbf{\Delta}\mathbf{y}_\mathcal{Q},
\label{equ_sys_math}
\end{equation}
where $\mathbf{\Delta}$ has the size of $M'NN_r\times MNN_r$. The
sum rate lower bound for the dynamic oversampled system is
\begin{equation}
C_{LB} = E\{\frac{1}{N}\log_2\det\left(\mathbf{\Delta}\mathbf{C}_{\mathbf{y}_\mathcal{Q}}\mathbf{\Delta}^H(\mathbf{\Delta}\mathbf{C}_{\mathbf{n}'}\mathbf{\Delta}^H)^{-1}\right)\}.
\label{equ_new_LB}
\end{equation}
The optimization problem that corresponds to the design of the
optimal $\mathbf{\Delta}^{\text{opt}}$ that can obtain the highest
achievable sum rate is described by
\begin{equation}
\mathbf{\Delta}^{\text{opt}} = \arg\max_\mathbf{\Delta}\enspace
\log_2\det\left(\mathbf{\Delta}\mathbf{C}_{\mathbf{y}_\mathcal{Q}}
\mathbf{\Delta}^H(\mathbf{\Delta}\mathbf{C}_{\mathbf{n}'}\mathbf{\Delta}^H)^{-1}\right).
\label{equ_optprob}
\end{equation}
Since the determinant is a log-concave function
\cite{boyd2004convex} and with the properties of the determinant,
(\ref{equ_optprob}) is simplified to
\begin{equation}
\mathbf{\Delta}^{\text{opt}} =  \arg\max_\mathbf{\Delta}\enspace
\frac{\det(\mathbf{\Delta}\mathbf{C}_{\mathbf{y}_\mathcal{Q}}
\mathbf{\Delta}^H)}{\det(\mathbf{\Delta}\mathbf{C}_{\mathbf{n}'}\mathbf{\Delta}^H)}.
\label{equ_optprob2}
\end{equation}
According to \cite{4270008}, the solution of (\ref{equ_optprob2}) is
equivalent to the solution of the following optimization problem
\begin{equation}
\mathbf{\Delta}^{\text{opt}} = \arg\max_\mathbf{\Delta}\enspace
\text{Tr}\left(\mathbf{\Delta}\mathbf{C}_{\mathbf{y}_\mathcal{Q}}
\mathbf{\Delta}^H(\mathbf{\Delta}\mathbf{C}_{\mathbf{n}'}\mathbf{\Delta}^H)^{-1}\right).
\label{equ_optprob3}
\end{equation}

\subsection{Mean Square Error based System Design}

In this system, the reduction matrix should be designed to obtain
the smallest MSE of the detected symbols. In order to reduce the
computational cost while performing detection, the sliding
window technique as discussed in \cite{8450809} is applied.

Similar to (\ref{equ_sys_math}), the dimension reduction operation
in each window can be mathematically described by
\begin{equation}
\mathbf{y}'_\mathcal{Q_{\text{win}}}=\mathbf{\Delta}_\text{win}\mathbf{y}_\mathcal{Q_{\text{win}}},
\label{equ_window}
\end{equation}
where $\mathbf{\Delta}_\text{win}$ is the reduction matrix with the
size of $M'lN_r\times MlN_r$ for each window and $l$ denotes the
window length.
$\mathbf{y}_\mathcal{Q_{\text{win}}}\in\mathbb{C}^{MlN_r\times 1}$
represents the received signal in each window. The optimization
problem that leads to the optimal linear detector is formulated as
\begin{equation}
\mathbf{W}_{\text{LRA-MMSE}} = \arg\min_{\mathbf{W}}\enspace E\left\{\left\vert\left\vert \mathbf{x}_{\text{win}}-\mathbf{W}^H\mathbf{y}'_{\mathcal{Q}_{\text{win}}}\right\vert\right\vert^2_2\right\},
\label{equ_lrammse}
\end{equation}
where $\mathbf{W}\in\mathbb{C}^{M'lN_r\times lN_t}$. The solution of
(\ref{equ_lrammse}) is given by
\begin{equation}
\begin{aligned}
\mathbf{W}_{\text{LRA-MMSE}} &= \mathbf{C}_{\mathbf{y}'_{\mathcal{Q}_{\text{win}}}}^{-1}\mathbf{C}_{\mathbf{y}'_{\mathcal{Q}_{\text{win}}}\mathbf{x}_{\text{win}}}\\&=(\mathbf{\Delta}_{\text{win}}\mathbf{C}_{\mathbf{y}_{\mathcal{Q}_{\text{win}}}}\mathbf{\Delta}_{\text{win}}^H)^{-1}\mathbf{\Delta}_{\text{win}}\mathbf{C}_{\mathbf{y}_{\mathcal{Q}_{\text{win}}}\mathbf{x}_{\text{win}}}.
\end{aligned}
\label{equ_LRAMMSE}
\end{equation}
According to (\ref{equ_Cyq}) and (\ref{equ_arcsin}), the covariance
matrix $\mathbf{C}_{\mathbf{y}_{\mathcal{Q}_{\text{win}}}}$ and the
cross-correlation matrix
$\mathbf{C}_{\mathbf{y}_{\mathcal{Q}_{\text{win}}}\mathbf{x}_{\text{win}}}$
are calculated by
\begin{equation}
\resizebox{\columnwidth}{!}{$\displaystyle
\mathbf{C}_{\mathbf{y}_{\mathcal{Q}_{\text{win}}}}=\frac{2}{\pi}\left(\text{sin}^{-1}(\mathbf{K}_{\text{win}}\mathbf{C}_{\mathbf{y}_{\text{win}}}^R\mathbf{K}_{\text{win}})+j\text{sin}^{-1}(\mathbf{K}_{\text{win}}\mathbf{C}_{\mathbf{y}_{\text{win}}}^I\mathbf{K}_{\text{win}})\right)$}
\end{equation}
\begin{equation}
\mathbf{C}_{\mathbf{y}_{\mathcal{Q}_{\text{win}}}\mathbf{x}_{\text{win}}} = \sqrt{\frac{2}{\pi}}\mathbf{K}_{\text{win}}\mathbf{H}_{\text{win}},\text{with}\enskip\mathbf{K}_{\text{win}} = \text{diag}(\mathbf{C}_{\mathbf{y}_{\text{win}}})^{-\frac{1}{2}},
\end{equation}
where $\mathbf{H}_{\text{win}}$ represents the equivalent channel
matrix in the window ($N=l$ in (\ref{equ:H})). The calculation of
$\mathbf{\Delta}_{\text{win}}$ leads to the optimization problem
\begin{equation}
\mathbf{\Delta}_{\text{win}}^\text{opt}=\arg\min_{\mathbf{\Delta}_{\text{win}}} E\left\{\left\vert\left\vert \mathbf{x}_{\text{win}}-\mathbf{W}^H_{\text{LRA-MMSE}}\mathbf{y}'_{\mathcal{Q}_{\text{win}}}\right\vert\right\vert^2_2\right\}.
\label{equ_optimization_delta}
\end{equation}
Inserting (\ref{equ_LRAMMSE}) into (\ref{equ_optimization_delta}) and expanding the terms, we obtain
\begin{equation}
\begin{aligned}
&\arg\min_{\mathbf{\Delta}_{\text{win}}} \text{Tr}\left(\mathbf{I}_{N_t}-(\mathbf{\Delta}_{\text{win}} \mathbf{C}_{\mathbf{y}_{\mathcal{Q}_{\text{win}}}\mathbf{x}_{\text{win}}})^H(\mathbf{\Delta}_{\text{win}}\mathbf{C}_{\mathbf{y}_{\mathcal{Q}_{\text{win}}}}\mathbf{\Delta}_{\text{win}}^{H})^{-1}\right.\\&\left.\qquad\qquad\qquad\qquad\qquad\qquad\qquad\qquad(\mathbf{\Delta}_{\text{win}}\mathbf{C}_{\mathbf{y}_{\mathcal{Q}}\mathbf{x}_{\text{win}}})\right),
\end{aligned}
\end{equation}
which can be further simplified to
\begin{equation}
\begin{aligned}
\mathbf{\Delta}_{\text{win}}^\text{opt}=&\arg\max_{\mathbf{\Delta}_{\text{win}}}
\text{Tr}\left(\mathbf{\Delta}_{\text{win}}\mathbf{C}_{\mathbf{y}_{\mathcal{Q}_{\text{win}}}
\mathbf{x}_{\text{win}}}\mathbf{C}_{\mathbf{y}_{\mathcal{Q}_{\text{win}}}\mathbf{x}_{\text{win}}}^H\right.\\&\left.\qquad\qquad\qquad
\mathbf{\Delta}_{\text{win}}^H(\mathbf{\Delta}_{\text{win}}
\mathbf{C}_{\mathbf{y}_{\mathcal{Q}_{\text{win}}}}\mathbf{\Delta}_{\text{win}}^{H})^{-1}\right)
\end{aligned}
\label{equ_new_opt}
\end{equation}
with
\begin{equation}
 \mathbf{C}_{\mathbf{y}_{\mathcal{Q}_{\text{win}}} \mathbf{x}_{\text{win}}}
 \mathbf{C}_{\mathbf{y}_{\mathcal{Q}_{\text{win}}}\mathbf{x}_{\text{win}}}^H =
 \frac{2}{\pi}\mathbf{K}_{\text{win}}\mathbf{H}_{\text{win}}\mathbf{H}_{\text{win}}^H\mathbf{K}_{\text{win}},
 \label{equ_cc}
\end{equation}
which is also a discrete optimization problem as the search for the
optimal sampling pattern using $\mathbf{\Delta}$ involves discrete
variables as the sampling instants. Surprisingly,
(\ref{equ_cc}) has the same solution with
$(\mathbf{C}_{\mathbf{y}_{\mathcal{Q}_{\text{win}}}}-\mathbf{C}_{\mathbf{n}'_{\text{win}}})$,
which is then inserted into (\ref{equ_new_opt}) and yields
    \begin{equation}
        \begin{aligned}
            \mathbf{\Delta}_{\text{win}}^\text{opt}=&\arg\max_{\mathbf{\Delta}_{\text{win}}}
            \text{Tr}\left(\mathbf{\Delta}_{\text{win}}(\mathbf{C}_{\mathbf{y}_{\mathcal{Q}_{\text{win}}}}-\mathbf{C}_{\mathbf{n}'_{\text{win}}})
            \mathbf{\Delta}_{\text{win}}^H(\mathbf{\Delta}_{\text{win}}
            \mathbf{C}_{\mathbf{y}_{\mathcal{Q}_{\text{win}}}}\mathbf{\Delta}_{\text{win}}^{H})^{-1}\right)\\=&\arg\max_{\mathbf{\Delta}_{\text{win}}}
            \text{Tr}\left(\mathbf{I}_{M'lN_r}-\mathbf{\Delta}_{\text{win}}\mathbf{C}_{\mathbf{n}'_{\text{win}}}
            \mathbf{\Delta}_{\text{win}}^H(\mathbf{\Delta}_{\text{win}}
            \mathbf{C}_{\mathbf{y}_{\mathcal{Q}_{\text{win}}}}\mathbf{\Delta}_{\text{win}}^{H})^{-1}\right)\\
            =&\arg\min_{\mathbf{\Delta}_{\text{win}}}
            \text{Tr}\left(\mathbf{\Delta}_{\text{win}}\mathbf{C}_{\mathbf{n}'_{\text{win}}}
            \mathbf{\Delta}_{\text{win}}^H(\mathbf{\Delta}_{\text{win}}
            \mathbf{C}_{\mathbf{y}_{\mathcal{Q}_{\text{win}}}}\mathbf{\Delta}_{\text{win}}^{H})^{-1}\right)\\=&\arg\max_{\mathbf{\Delta}_{\text{win}}}
            \text{Tr}\left(\mathbf{\Delta}_{\text{win}}\mathbf{C}_{\mathbf{y}_{\mathcal{Q}_{\text{win}}}}
            \mathbf{\Delta}_{\text{win}}^H(\mathbf{\Delta}_{\text{win}}\mathbf{C}_{\mathbf{n}'_{\text{win}}}
            \mathbf{\Delta}_{\text{win}}^{H})^{-1}\right).
        \end{aligned}
        \label{equ_new_opt2}
    \end{equation}
    Note that the optimization problems in \eqref{equ_optprob3} and \eqref{equ_new_opt2} have similar objective functions. Especially when $\mathbf{\Delta}^\text{opt}$ has the same matrix size with $ \mathbf{\Delta}_{\text{win}}^\text{opt}$, the solutions are equal. In other words, the optimal solution $\mathbf{\Delta}^\text{opt}$ in \eqref{equ_optprob3} (or $ \mathbf{\Delta}_{\text{win}}^\text{opt}$ in \eqref{equ_new_opt2}) will both maximize the sum rate and minimize the MSE of detected symbols.

After obtaining the optimal reduction matrix
$\mathbf{\Delta}_{\text{win}}^{\text{opt}}$, the sliding-window
based LRA-MMSE detector for the dynamic oversampled systems is then
given by
\begin{equation}
\mathbf{W}_{\text{LRA-MMSE}}^\text{opt} =
(\mathbf{\Delta}_{\text{win}}^{\text{opt}}
\mathbf{C}_{\mathbf{y}_{\mathcal{Q}_{\text{win}}}}{\mathbf{\Delta}_{\text{win}}^{\text{opt}}}^H)^{-1}
\mathbf{\Delta}_{\text{win}}^{\text{opt}}\mathbf{C}_{\mathbf{y}_{\mathcal{Q}_{\text{win}}}\mathbf{x}_{\text{win}}}
\label{equ_MMSE_detector}
\end{equation}
and the detected symbols in one window are obtained through
\begin{equation}
\tilde{\mathbf{x}}_{\text{win}}=\mathbf{W}_{\text{LRA-MMSE}}^{\text{opt}^H}\mathbf{\Delta}_\text{win}^\text{opt}\mathbf{y}_\mathcal{Q_{\text{win}}}.
\label{equ_detected_symbol}
\end{equation}

\section{Design of the Dimension Reduction Matrix}

In this section, design algorithms are presented for the reduction
matrix $\mathbf{\Delta}$ that operates on the oversampled signal and
performs dynamic oversampling by combining or choosing the samples
according to the sum rate or the MSE criteria. In particular,
algorithms are presented to solve the optimization problems
(\ref{equ_optprob3}) and (\ref{equ_new_opt}), both of which can be
generalized as
\begin{equation}
\mathbf{\Delta}^{\text{opt}} = \arg\max_\mathbf{\Delta}\enspace \text{Tr}\left(\mathbf{\Delta}\mathbf{C}_A\mathbf{\Delta}^H(\mathbf{\Delta}\mathbf{C}_B\mathbf{\Delta}^H)^{-1}\right),
\label{equ_gen_opt}
\end{equation}
where both $\mathbf{C}_A$ and $\mathbf{C}_B$ have the size of
$MNN_r\times MNN_r$ and $\mathbf{\Delta}$ has the size of
$M'NN_r\times MNN_r$. The problem in (\ref{equ_gen_opt}) is known as
the ratio trace problem \cite{4801520}. The GEVD is first considered
to obtain a full dimension reduction matrix with complex
coefficients, which is shown to have a relatively high computational
cost. Then, inspired by signal sampling principles, the
design of sparse binary form of $\mathbf{\Delta}$ is proposed. It performs
sampling using ones in the positions where the sample is chosen and
zeros for the discarded samples. In addition, we devise strategies
for designing sparse binary matrices based on a backward feature selection
algorithm and a novel greedy approach that employs sequential search.

\subsection{Generalized Eigenvalue Decomposition}

From \cite{4801520}, the problem in (\ref{equ_gen_opt}) can be
solved by the GEVD described by
\begin{equation}
    \mathbf{C}_A\mathbf{\delta}_c=\lambda_c\mathbf{C}_B\mathbf{\delta}_c,
\end{equation}
where $\lambda_c$ is the $c$th largest generalized eigenvalue. The
matrix ${\mathbf{\Delta}^{\text{opt}}}^H$ is then constituted by the
corresponding eigenvectors $\mathbf{\delta}_c$, $c=1,\cdots,M'NN_r$.
A step-by-step GEVD solution \cite{ghojogh2019eigenvalue} is
summarized in Alg. \ref{alg:receiver}. In step 5, $\mathbf{\Lambda}$
is a diagonal matrix containing all the eigenvalues and the
corresponding eigenvectors are stored in $\mathbf{\Delta}$.

\begin{algorithm}
    \caption{GEVD}
    \begin{algorithmic}[1]
        \State Eigenvalue decomposition: $\mathbf{\Phi}_{\text{B}},\mathbf{\Lambda}_{\text{B}}\leftarrow\mathbf{C}_{\mathbf{B}}\mathbf{\Phi}_{\text{B}}=\mathbf{\Phi}_{\text{B}}\mathbf{\Lambda}_{\text{B}}$
        \State $\tilde{\mathbf{\Phi}}_{\text{B}}\leftarrow\tilde{\mathbf{\Phi}}_{\text{B}}=\mathbf{\Phi}_{\text{B}}\mathbf{\Lambda}^{-\frac{1}{2}}_{\text{B}}$
        \State $\mathbf{A}\leftarrow\mathbf{A}=\tilde{\mathbf{\Phi}}^H_{\text{B}}\mathbf{C}_{\mathbf{A}}\tilde{\mathbf{\Phi}}_{\text{B}}$
        \State Eigenvalue decomposition:  $\mathbf{\Phi}_\text{A},\mathbf{\Lambda}_\text{A}\leftarrow\mathbf{A}\mathbf{\Phi}_\text{A}=\mathbf{\Phi}_\text{A}\mathbf{\Lambda}_\text{A}$
        \State $\mathbf{\Lambda}\leftarrow\mathbf{\Lambda}_\text{A}$
        \State $\mathbf{\Delta}\leftarrow\mathbf{\Delta}=\tilde{\mathbf{\Phi}}_{\text{B}}\mathbf{\Phi}_{\text{A}}$
        \State Extract the eigenvectors in $\mathbf{\Delta}$ related to the $M'NN_{r}$ largest diagonal values in $\mathbf{\Lambda}$
        \State Build ${\mathbf{\Delta}^\text{opt}}^H$ column by column
    \end{algorithmic}
    \label{alg:receiver}
\end{algorithm}

\subsection{Submatrix-level Feature Selection}
\label{sec_submatrix}

The main drawback of the GEVD algorithm is the high computational
cost of the eigenvalue decompositions and matrix multiplications with full matrices involved in (\ref{equ_sys_math}) and (\ref{equ_window}). Considering this, we devise an approach, in which the dimension reduction matrix
is a sparse binary matrix that contains a single one in each row.
The advantage of this sparse binary matrix $\mathbf{\Delta}_B$ is
that when multiplying such matrix by a received vector only a few
data samples are selected. In addition, the samples associated with
zeros in $\mathbf{\Delta}_B$ are discarded without the need for
arithmetic operations, which can largely reduce the computational
cost. The optimization problem in (\ref{equ_gen_opt}) can be
reformulated as
\begin{equation}
\mathbf{\Delta}^{\text{opt}}_B = \underset{\text{$\mathbf{\Delta}_B$ is sparse binary}}{\arg\max}\enspace\text{Tr}\left(\mathbf{\Delta}_B\mathbf{C}_A\mathbf{\Delta}_B^H(\mathbf{\Delta}_B\mathbf{C}_B\mathbf{\Delta}_B^H)^{-1}\right),
\label{equ_binopt}
\end{equation}
    which is a discrete optimization problem because the
    search for the optimal sampling pattern involves discrete variables as the sampling instants. Specifically, the dimension reduction matrix is constrained to have only ones and zeros and performs selection of the samples. This constraint on the reduction matrix turns its design into a discrete optimization problem in which we must determine the positions of the ones over a range of discrete values.

However, (\ref{equ_binopt}) is not easy to solve due to its
combinatorial nature. The optimal but most costly way is to search
for all possible patterns of $\mathbf{\Delta}_B$ and select the best
one. In large-scale MIMO systems, $\mathbf{\Delta}_B$ has large
dimensions, which result in a high computational cost while
conducting the search. In order to alleviate this cost, by
exploiting the sparse nature of $\mathbf{\Delta}_B$ the search is
split over several low dimensional submatrices given by
\begin{equation}
\mathbf{\Delta}_B = \text{blkdiag}
([{\mathbf{\Delta}_B}_1, {\mathbf{\Delta}_B}_2, \cdots, {\mathbf{\Delta}_B}_K])
\label{equ_deltab}
\end{equation}
and search for the best pattern for each submatrix
${\mathbf{\Delta}_B}_k$ $(k=1,\cdots,K)$ with the size of
$\frac{M'NN_r}{K}\times \frac{MNN_r}{K}$. The choice of $K$ depends
on the value of $M'NN_r$, so that $\frac{M'NN_r}{K}$ cannot be very
small nor large. Since a very low dimensional matrix
${\mathbf{\Delta}_B}_k$ cannot accurately represent the original
matrix ${\mathbf{\Delta}_B}$ and a high dimensional matrix will turn
the search costly, there is a trade-off. The
optimization problem is then reduced to
\begin{equation}
\resizebox{.97\columnwidth}{!}{$\displaystyle
\mathbf{\Delta}_{B_k}^{\text{opt}} =
\underset{\text{$\mathbf{\Delta}_{B_k}$ is sparse
binary}}{\arg\max}\enspace\text{Tr}\left(\mathbf{\Delta}_{B_k}\mathbf{C}_{A_k}
\mathbf{\Delta}_{B_k}^H(\mathbf{\Delta}_{B_k}\mathbf{C}_{B_k}\mathbf{\Delta}_{B_k}^H)^{-1}\right)$},
\label{equ_binopt2}
\end{equation}
where $\mathbf{C}_{A_k},
\mathbf{C}_{B_k}\in\mathbb{C}^{\frac{MNN_r}{K}\times
\frac{MNN_r}{K}}$ are block diagonal submatrices from
$\mathbf{C}_{A}$ and $\mathbf{C}_{B}$, respectively. In the
following, two algorithms are illustrated for searching the optimal
pattern $\mathbf{\Delta}_{B_k}^{\text{opt}}$, submatrix-level
backward feature selection (SL-BFS) and submatrix-level restricted
greedy search (SL-RGS). A simplified SL-FS algorithm is then
proposed to reduce the search cost.

\subsubsection{Backward Feature Selection}

Inspired by the feature selection algorithms used in machine
learning and statistics \cite{devyver1982pattern}, the
idea of BFS is extended to search for $\mathbf{\Delta}_{B_k}^{\text{opt}}$ in
the sub-matrix level, which is called the SL-BFS algorithm. The initialization is an identity matrix and the least significant row is removed at each
iteration. The proposed SL-BFS algorithm attempts to obtain the most
suitable positions for the one and the zeros in each row of the
submatrix according to the criterion in (\ref{equ_binopt2}). At the
end of the iterations, the rows contributing to the smallest trace
in (\ref{equ_binopt2}) are eliminated. The details of the BFS
algorithm are summarized in Alg. \ref{alg:receiver2}.

\begin{algorithm}
    \caption{SL-BFS}
    \begin{algorithmic}[1]
        \State \textbf{\underline{Input}:}\quad$\mathbf{C}_{A_k}, \mathbf{C}_{B_k}$
        \State \textbf{\underline{Output}:}\quad$\mathbf{\Delta}_{B_k}^{\text{opt}}$
        \State \textbf{\underline{Algorithm}:}
        \State $\mathbf{\Delta}_{B_k}^{\text{opt}}=\mathbf{I}_{\frac{MNN_r}{K}}$
        \For{$r=1:\frac{(M-M')NN_r}{K}$}
        \State $\mathbf{\Delta}_{B_k}^{\text{tmp1}}=\mathbf{\Delta}_{B_k}^{\text{opt}}$, $\mathbf{\Delta}_{B_k}^{\text{tmp2}}=\mathbf{\Delta}_{B_k}^{\text{opt}}$, $s_{k_{\max}}=0$
        \For{$rr=1:\frac{MNN_r}{K}-r+1$}
        \State Delete the $rr$th row of $\mathbf{\Delta}_{B_k}^{\text{tmp1}}$
        \State $s_k=\text{Tr}\left({\mathbf{\Delta}_{B_k}^{\text{tmp1}}}\mathbf{C}_{A_k}{\mathbf{\Delta}_{B_k}^{\text{tmp1}}}^H(\mathbf{\Delta}_{B_k}^{\text{tmp1}}\mathbf{C}_{B_k}{\mathbf{\Delta}_{B_k}^{\text{tmp1}}}^H)^{-1}\right)$
        \If{$s_k > s_{k_{\max}}$}
        \State $s_{k_{\max}}=s_k$
        \State $\mathbf{\Delta}_{B_k}^{\text{opt}} = \mathbf{\Delta}_{B_k}^{\text{tmp1}}$
        \EndIf
        \State $\mathbf{\Delta}_{B_k}^{\text{tmp1}} = \mathbf{\Delta}_{B_k}^{\text{tmp2}}$
        \EndFor
        \EndFor
    \end{algorithmic}
    \label{alg:receiver2}
\end{algorithm}

\subsubsection{Restricted Greedy Search}

The basic idea of the proposed SL-RGS algorithm is that based on the
initialized pattern $\mathbf{\Delta}_{B_k}^I$ the best row patterns
are sequentially searched from the first until the last row. While
searching for the $r$th row pattern in the submatrix, the
position of the one is shifted within a pre-defined small range and the
one contributing to the larges trace in (\ref{equ_binopt2}) is selected. In the
computation of the trace, the remaining $\frac{M'NN_r}{K}-1$ rows
are fixed including the first $r-1$ optimized rows and the remaining
$\frac{M'NN_r}{K}-r$ non-optimized rows. The proposed SL-RGS
algorithm is described in Alg. \ref{alg:receiver3}.

\begin{algorithm}
    \caption{SL-RGS}
    \begin{algorithmic}[1]
        \State \textbf{\underline{Input}:}\quad$\mathbf{C}_{A_k}, \mathbf{C}_{B_k}, \mathbf{\Delta}_{B_k}^I$
        \State \textbf{\underline{Output}:}\quad$\mathbf{\Delta}_{B_k}^{\text{opt}}$
        \State \textbf{\underline{Algorithm}:}
        \State $\mathbf{\Delta}_{B_k}^{\text{tmp}}=\mathbf{\Delta}_{B_k}^I$, $\mathbf{\Delta}_{B_k}^{\text{opt}}=\mathbf{\Delta}_{B_k}^I$
        \State Find the position of the one in each row of $\mathbf{\Delta}_{B_k}^I$ and store them into vector $\mathbf{j}$
        \State $s_{\max_k}=\text{Tr}\left({\mathbf{\Delta}_{B_k}^I}\mathbf{C}_{A_k}\mathbf{\Delta}_{B_k}^{I^H}(\mathbf{\Delta}_{B_k}^I\mathbf{C}_{B_k}\mathbf{\Delta}_{B_k}^{I^H})^{-1}\right)$
        \For{$r=1:\frac{M'NN_r}{K}$}
        \State Set the position $(r,j_r)$ in $\mathbf{\Delta}_{B_k}^{\text{tmp}}$ to zero
        \For{$c=j_r-\beta:j_r+\beta$}
        \If{$c\notin\mathbf{j}$}
        \State Set the position $(r,c)$ in $\mathbf{\Delta}_{B_k}^{\text{tmp}}$ to one
        \State $s=\text{Tr}\left({\mathbf{\Delta}_{B_k}^{\text{tmp}}}\mathbf{C}_{A_k}{\mathbf{\Delta}_{B_k}^{\text{tmp}}}^H(\mathbf{\Delta}_{B_k}^{\text{tmp}}\mathbf{C}_{B_k}{\mathbf{\Delta}_{B_k}^{\text{tmp}}}^H)^{-1}\right)$
        \If{$s>s_{\max_k}$}
        \State $s_{\max_k}=s$
        \State $\mathbf{\Delta}_{B_k}^{\text{opt}} = \mathbf{\Delta}_{B_k}^{\text{tmp}}$
        \EndIf
        \State Set the position $(r,c)$ in $\mathbf{\Delta}_{B_k}^{\text{tmp}}$ to zero
        \EndIf
        \EndFor
        \State $\mathbf{\Delta}_{B_k}^{\text{tmp}} = \mathbf{\Delta}_{B_k}^{\text{opt}}$
        \State Update $\mathbf{j}$
        \EndFor
    \end{algorithmic}
    \label{alg:receiver3}
\end{algorithm}
The initialized pattern $\mathbf{\Delta}_{B_k}^I$ is the $k$th block
diagonal submatrix of $\mathbf{\Delta}_{B}^{I}$ (as in
(\ref{equ_deltab})). The matrix $\mathbf{\Delta}_{B}^{I}$ is the
initialized pattern for $\mathbf{\Delta}_{B}$ and is selected as the
pattern for uniform or quasi-uniform\footnote{When $\frac{M}{M'}$ is
an integer, $\mathbf{\Delta}_{B}^{I}$ is the pattern for
uniform oversampling, otherwise it is for quasi-uniform
oversampling.} oversampling, i.e.,

\begin{equation}
\mathbf{\Delta}_B^I=\begin{bmatrix}
\undermat{r_1}{0 & \cdots & 0} & 1 & 0 & 0 & 0 & 0 & \cdots & 0 & 0 & 0 &\\
\vdots &  & \vdots & \vdots & \vdots & \vdots & \vdots & \vdots & \vdots & \vdots & \vdots & \vdots &\\
\undermat{r_r}{0 & \cdots & 0 & 0} & 1 & 0 & 0 & 0 & \cdots & 0 & 0 & 0 &\\
\vdots &  &  & \vdots & \vdots & \vdots & \vdots & \vdots & \vdots & \vdots & \vdots & \vdots &\\
\undermat{r_{\text{end}}}{0 & \cdots & 0 & 0 & 0 & 0 & 0} & 1 & \cdots & 0 & 0 & 0 &\\
\end{bmatrix},
\label{equ_pattern}
\end{equation}
\vspace{0.85em}

where $r_r$ denotes the number of zeros before 1 in each row. In the
following, several examples are taken to illustrate how
$\mathbf{\Delta}_B^I$ is chosen. For a system with $M=4$ (or $M=6$)
and $M'=2$, $\mathbf{\Delta}_B^I$ can be easily found. Since the
uniform sampling pattern in one Nyquist interval is [1,0,1,0] (or
[1,0,0,1,0,0] for $M=6$), $r_r$ is set as $M\frac{r-1}{M'}$, where
$r$ is the index of the row. However, for the system with $M=3$ (or
$M=5$) and $M'=2$, the quasi-uniform sampling pattern in one Nyquist
interval has more alternatives, which can be either [1,0,1] or
[1,1,0] (either [1,0,1,0,0] or [1,0,0,1,0] for $M=5$). $r_r$ is then
set as
\begin{equation*}
r_r=\begin{cases}
M\lfloor\frac{r-1}{M'}\rfloor, & \text{if $\mod(r-1,M')=0$}\\
M\lfloor\frac{r-1}{M'}\rfloor+\alpha, & \text{otherwise},
\end{cases}
\end{equation*}
where $\alpha$ denotes the position of the second one starting from position 0.

In step 8, $j_r$ denotes the position of the one at the $r$th row
and is stored in the vector $\mathbf{j}\in\mathbb{R}^{1\times
\frac{M'NN_r}{K}}$. The parameter $\beta$ in step 9 is the
pre-defined range number, which aims to reduce the cost of the
search. The possible position of the one is only searched within the
range of $2\beta$ around $j_r$.

\subsubsection{Simplified SL-FS}

Since the computation of $\mathbf{\Delta}_{B}^\text{opt}$ is costly,
parts of the SL-FS in both SL-BFS and SL-RGS algorithms need to be
repeated $K$ times in order to obtain all optimal submatrices
$\mathbf{\Delta}_{B_k}^\text{opt}$ ($k=1,\cdots,K$). To further
reduce the complexity of both SL-BFS and SL-RGS algorithms, we
propose simplified versions of SL-FS in which all optimal
submatrices $\mathbf{\Delta}_{B_k}^\text{opt}$ ($k=1,\cdots,K$) are
assumed to share the same pattern so that only one submatrix
$\mathbf{\Delta}_{B_k}$ will be optimized by the SL-FS recursions
detailed in Alg. \ref{alg:receiver4}. The choice of which
$\mathbf{\Delta}_{B_k}$ is optimized is influenced by the smallest
trace calculated by its initial pattern, which is detailed from
steps 1 to 4 in Alg. \ref{alg:receiver4}.

\begin{algorithm}
    \caption{Simplified SL-FS}
    \begin{algorithmic}[1]
        \For{$k=1:K$}
        \State $s_{k}=\text{Tr}\left({\mathbf{\Delta}_{B_k}^{I}}\mathbf{C}_{A_k}\mathbf{\Delta}_{B_k}^{I^H}(\mathbf{\Delta}_{B_k}^{I}\mathbf{C}_{B_k}\mathbf{\Delta}_{B_k}^{I^H})^{-1}\right)$
        \EndFor
        \State Choose the $\mathbf{\Delta}_{B_k}^{I}$ with the lowest $s_{k}$
        \State Use the SL-BFS or SL-RGS algorithm to find $\mathbf{\Delta}_{B_k}^{\text{opt}}$
        \State Let $\mathbf{\Delta}_{B_1}^{\text{opt}}=\cdots=\mathbf{\Delta}_{B_k}^{\text{opt}}=\cdots=\mathbf{\Delta}_{B_K}^{\text{opt}}$
        \State Rebuild $\mathbf{\Delta}_B^{\text{opt}}$ as (\ref{equ_deltab})
    \end{algorithmic}
    \label{alg:receiver4}
\end{algorithm}

\section{Analysis of Proposed Scheme}

In this section, we study the convergence of the proposed SL-RGS
algorithm, compare the computational complexities of different
reduction algorithms, analyze the performance impacts of various $K$ and $\beta$ and illustrate the power consumption of the
proposed scheme at the receiver.

\subsection{Convergence of the SL-RGS Algorithm}

We focus on the proposed SL-RGS algorithm because the GEVD is an
established technique and the proposed SL-BFS is an extension of the
existing BFS, which has been studied in \cite{kittler1978feature}.
We remark that the study carried out here is not a theoretically
rigorous analysis of the convergence of the proposed SL-RGS
algorithm. In fact, it is a study that provides some insights on how
the proposed SL-RGS algorithm works and how it converges to a
cost-effective dimension reduction matrix with the help of
simulations.

Considering one sparse dimension reduction submatrix
$\mathbf{\Delta}_{B_k}$ defined in (\ref{equ_binopt2}), we would
like to examine the convergence of the proposed SL-RGS algorithm to
the unrestricted (or exhaustive search-based) GS algorithm under
some reasonable assumptions. Note that the exhaustive search for
$\mathbf{\Delta}_{B_k}^{\rm opt}$ results in
\scalebox{0.9}{$\left(\frac{(M-M')NN_r}{K}\right)^{\frac{M'NN_r}{K}}$}
independent calculations of the trace (step 12 in Alg.
\ref{alg:receiver3}), which is quite complex for practical use.

The proposed SL-RGS algorithm greatly reduces the cost of the search
(maximum \scalebox{0.95}{$\left(2\beta\right)^{\frac{M'NN_r}{K}}$}
independent calculations of trace) and under some conditions can
approach the performance of the unrestricted GS obtained
$\mathbf{\Delta}_{B_k}^{\rm opt}$. Assume that the SL-RGS
examines a range of patterns characterized by
\begin{equation}
\resizebox{\columnwidth}{!}{$\displaystyle
\mathbf{\Delta}_{B_k}=\begin{bmatrix} \undermat{\mathbf{c}_1}{0 &
    \ldots & 0 & 1 & 0 & \ldots & 0} & 0 & 0 &  0 & \ldots & 0
\\ \\
0 & \ldots & 0 & \undermat{\mathbf{c}_2}{0 & \ldots & 0 & 1 & 0 &
    \ldots & 0} & 0 & \cdots & 0 \\ \\
\vdots &  \vdots & \vdots  & \vdots & \vdots & \vdots & \vdots &
\vdots & \vdots & \vdots & \vdots & \vdots & \vdots \\ \\
0 & \ldots & 0 & 0 & 0 & 0 &\undermat{\mathbf{c}_{\frac{M'NN_r}{K}}}{0 & \ldots & 0 & 1 & 0 & \ldots & 0}
\\\\
\end{bmatrix}$},
\label{equ_pattern3}
\end{equation}
where $\mathbf{c}_r = [ 0  \ldots 0 ~ 1 ~ 0  \ldots  0]$
($r=1,2,\ldots,\frac{M'NN_r}{K}$) is a row vector that has length
$2\beta+1$. The trace of the $r$th optimized
$\mathbf{\Delta}_{B_k}^r$ is calculated as
\begin{equation}
s_r =
\text{Tr}\left(\mathbf{\Delta}_{B_k}^r\mathbf{C}_{A_k}{\mathbf{\Delta}_{B_k}^r}^H(\mathbf{\Delta}_{B_k}^r\mathbf{C}_{B_k}{\mathbf{\Delta}_{B_k}^r}^H)^{-1}\right).
\label{s_trace}
\end{equation}

\begin{figure}[!htbp]
    \centering
    \includegraphics{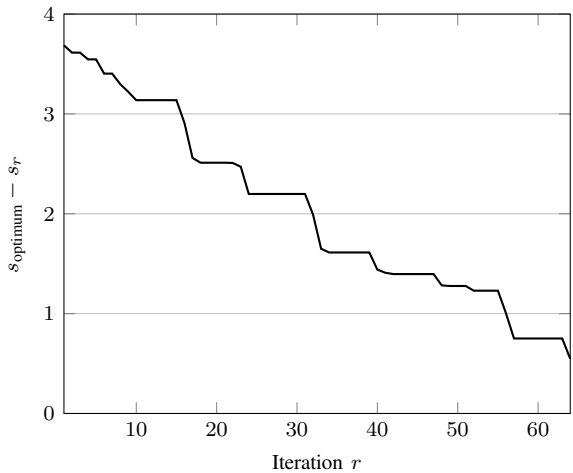}
    \caption{Convergence of the proposed SL-RGS to the unrestricted GS algorithm with $M'=2$, $N=4$, $N_r=64$ and $K=8$.}
    \label{fig:convergence}
\end{figure}

In the following, examples will be taken to illustrate how the
SL-RGS algorithm leads to
\begin{equation}
s_r \rightarrow s_{\rm optimum},
\end{equation}
where $s_{\rm optimum}$ is the trace obtained by the unrestricted GS with the form as
\begin{equation}
\resizebox{\columnwidth}{!}{$\displaystyle
\mathbf{\Delta}_{B_k}^{\rm optimum} =\begin{bmatrix} 0 & 1 & 0 & 0 & 0 & 0
& 0 & 0 & 0
& 0 &  0 & \ldots & 0 \\ \\
0 & 0 & 0 & 0 & 0 & 0 & 0 & 0 & 1 & 0 & 0 & \ldots & 0 \\ \\
0 & \ldots & 0 & 0 & 1 & 0 & 0 & 0 & 0 & 0 & 0 & 0 & 0
\\\\
\end{bmatrix}$}.
\label{equ_pattern4}
\end{equation}
In the SL-RGS algorithm, the initialized pattern
$\mathbf{\Delta}_{B_k}^{I}$ is
\begin{equation}
\mathbf{\Delta}_{B_k}^{I} =\begin{bmatrix} 1 & 0 & 0 & 0 & 0 & 0
& 0 & 0 & 0
& 0 &  0 & \ldots & 0 \\ \\
0 & 0 & 1 & 0 & 0 & 0 & 0 & 0 & 0 & 0 & 0 & \ldots & 0 \\ \\
0 & \ldots & 0 & 0 & 0 & 0 & 0 & 0 & 0 & 0 & 0 & 0 & 1
\\\\
\end{bmatrix}.
\label{equ_pattern5}
\end{equation}
After the 1st iteration, $\mathbf{\Delta}_{B_k}^{I}$ is changed to
\begin{equation}
\mathbf{\Delta}_{B_k}^{1} =\begin{bmatrix} 0 & 0 & 0 & 1 & 0 & 0
& 0 & 0 & 0
& 0 &  0 & \ldots & 0 \\ \\
0 & 0 & 1 & 0 & 0 & 0 & 0 & 0 & 0 & 0 & 0 & \ldots & 0 \\ \\
0 & \ldots & 0 & 0 & 0 & 0 & 0 & 0 & 0 & 0 & 0 & 0 & 1
\\\\
\end{bmatrix}.
\label{equ_pattern6}
\end{equation}
After the $\frac{M'NN_r}{K}$th iteration, the optimized submatrix is
\begin{equation}
\resizebox{\columnwidth}{!}{$\displaystyle
\mathbf{\Delta}_{B_{k}}^{\frac{M'NN_r}{K}} =\begin{bmatrix} 0 & 0 & 0 & 1 & 0 & 0
& 0 & 0 & 0
& 0 &  0 & \ldots & 0 \\ \\
0 & 0 & 0 & 0 & 0 & 1 & 0 & 0 & 0 & 0 & 0 & \ldots & 0 \\ \\
0 & \ldots & 0 & 0 & 0 & 0 & 0 & 0 & 0 & 1 & 0 & 0 & 0
\\\\
\end{bmatrix}$}.
\label{equ_pattern2}
\end{equation}
Fig. \ref{fig:convergence} shows the corresponding convergence
performance, where $r=1,2,\cdots,\frac{M'NN_r}{K}$. It can be seen that although
$\mathbf{\Delta}_{B_{k}}^{\frac{M'NN_r}{K}}$ and
$\mathbf{\Delta}_{B_k}^{\rm optimum}$ do not have similar patterns
the difference of trace $(s_{\text{optimum}}-s_r)$ still decreases as the
number of iterations increases until it approaches zero.
This indicates that the proposed SL-RGS algorithm leads to a close
to optimal result in terms of $s_r$, i.e. $s_r\rightarrow s_{\rm
optimum}$ provided that the range $2\beta$ is sufficiently large.
This discussion aims to give insight on how the proposed SL-RGS
algorithm can obtain satisfactory results provided the range $2
\beta$ is chosen sufficiently large.

\subsection{Computational Complexity}
\label{sec_complexity}

\subsubsection{GEVD vs SL-FS}

In the GEVD algorithm, the operations that contribute the most to
the computational complexity lie in the eigenvalue decomposition and
matrix multiplications. While in the SL-FS algorithms, the operation
of trace in each iteration consumes most of the calculations. The
computational complexities of the illustrated design algorithms are
shown in Table \ref{tab_com}, where $\mathcal{O}(\cdot)$ represents
the big O notation.

\begin{table*}[]
    \centering
    \caption{Computational complexity of the illustrated matrix design algorithms}
    \includegraphics{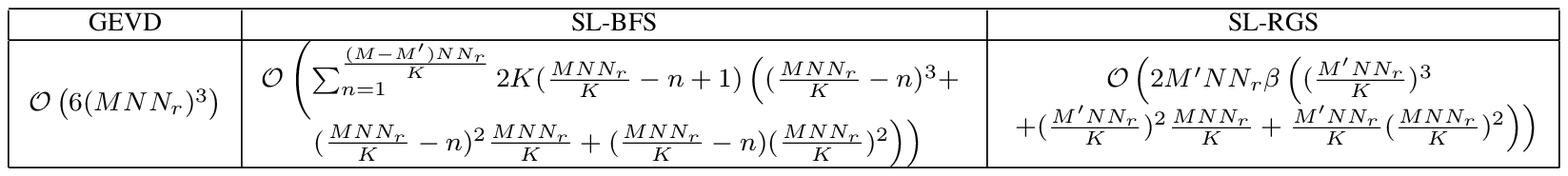}
    \label{tab_com}
\end{table*}

\begin{figure}[!htbp]
    \centering
    \includegraphics{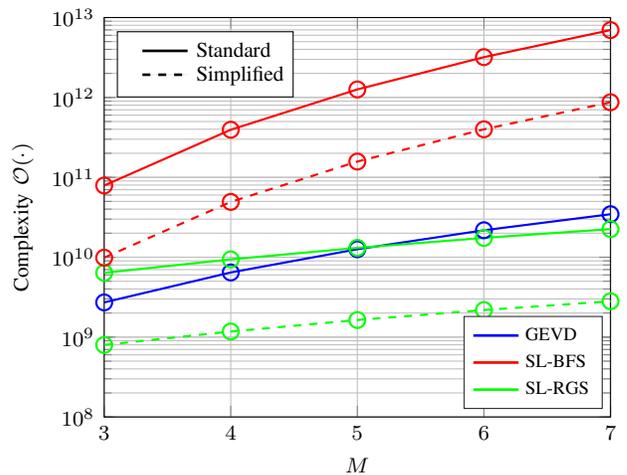}
    \caption{Computational complexity of analyzed matrix design algorithms as a function of initial sampling rate, which is $M$ times the Nyquist rate, where $N=4, N_r=64, M'=2, \beta=5$ and $K=8$}
    \label{fig:complexity}
\end{figure}

The solid lines in Fig. \ref{fig:complexity} represent the
computational complexity of the three matrix design algorithms
studied. It is noticed that the SL-RGS algorithm consumes roughly the
same complexity as the GEVD and the SL-BFS algorithm has the highest
complexity. The complexity of the simplified SL-BFS and SL-RGS are
shown as the dashed lines in Fig. \ref{fig:complexity}. Compared to
the standard SL-BFS and SL-RGS algorithms, the simplified SL-FS
recursions can provide savings of up to 87.5\% of the computational
costs.

\subsubsection{Uniform vs Dynamic Oversampling}

The comparison of the computational complexity of
different oversampling techniques using the MSE based design is
examined here. We remark that the window technique is considered for
both uniform and dynamic oversampling schemes. In systems with
uniform oversampling, all received samples are used for signal
processing, which means that there is no need for the pattern design
and dimension reduction. The corresponding sliding window based
LRA-MMSE detector is a modified version of
(\ref{equ_MMSE_detector}), which is described by
\begin{equation}
\mathbf{W}_{\text{LRA-MMSE}}^{\text{uni}} =
\mathbf{C}_{\mathbf{y}_{\mathcal{Q}_{\text{win}}}}^{-1}
\mathbf{C}_{\mathbf{y}_{\mathcal{Q}_{\text{win}}}\mathbf{x}_{\text{win}}}.
\end{equation}

\begin{figure}[!htbp]
    \centering
    \includegraphics{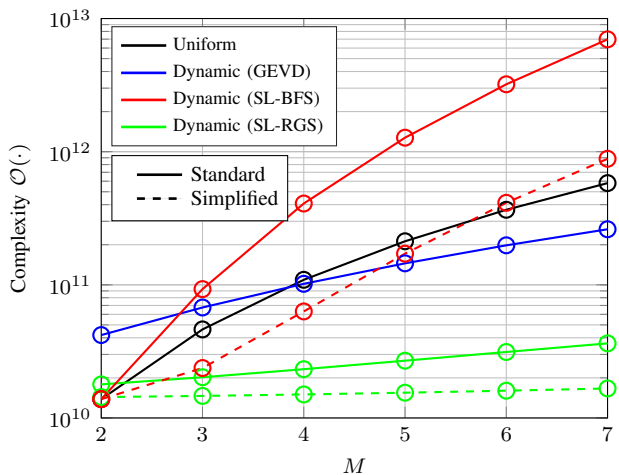}
    \caption{Detection complexity of uniform ($M=M'$) and dynamic oversampling in one
    transmission block (containing 100 symbols) as a function of initial sampling rate,
    which is $M$ times the Nyquist rate, where $l=4, N_r=64, N_t=4, M'=2, \beta=5$ and $K=8$.}
    \label{fig:complexity2}
\end{figure}

Table \ref{tab_com2} shows the computational complexity of different
oversampling techniques for obtaining the detected symbols in each
window with the known pattern design. Assuming that the whole
transmission block including the operation of pattern design, which
is obtained only once at the beginning of each transmission, and
dimension reduction, which is applied in each window, in systems
with dynamic oversampling the total complexity comparison between
uniform and dynamic oversampling techniques are shown in Fig.
\ref{fig:complexity2}. It can be noticed that both the standard and
simplified SL-RGS based dynamic oversampling techniques require the
lowest computational cost among the studied oversampling techniques.
We remark that the complexity reduction for dynamic oversampling
comes from the fact that a lower processing rate $M'=2$ is used
regardless of the sampling rate $M$, while the black curve from
uniform sampling increases the processing rate together with the
sampling rate $M'=M$. Compared to the standard algorithm, the
simplified SL-BFS and SL-RGS have saved up to 87.33\% and 54.16\% in
terms of computational costs, respectively. When compared with
uniform oversampling, dynamic oversampling with the GEVD and the
simplified SL-RGS techniques can save up to 54.99\% and 97.13\% in
terms of computational costs, respectively.

\begin{table*}[]
    \centering
    \caption{Computational complexity for obtaining the detected symbols in each window}
    \includegraphics{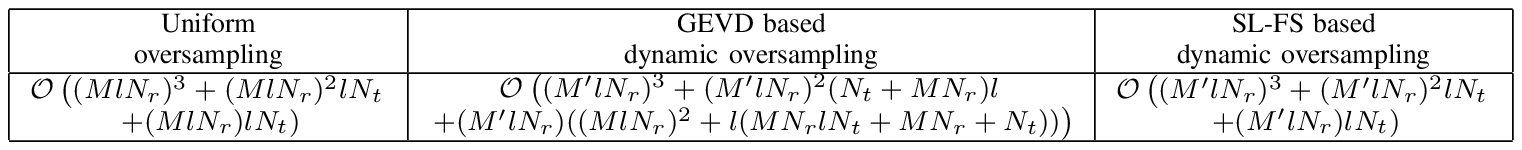}
    \label{tab_com2}
\end{table*}

\subsection{The choice of $K$ and $\beta$}

\begin{figure}[!htbp]
    \centering
    \subfigure[$\beta=5$]{
    \begin{minipage}[t]{0.5\linewidth}
        \centering
        \includegraphics[width=\columnwidth]{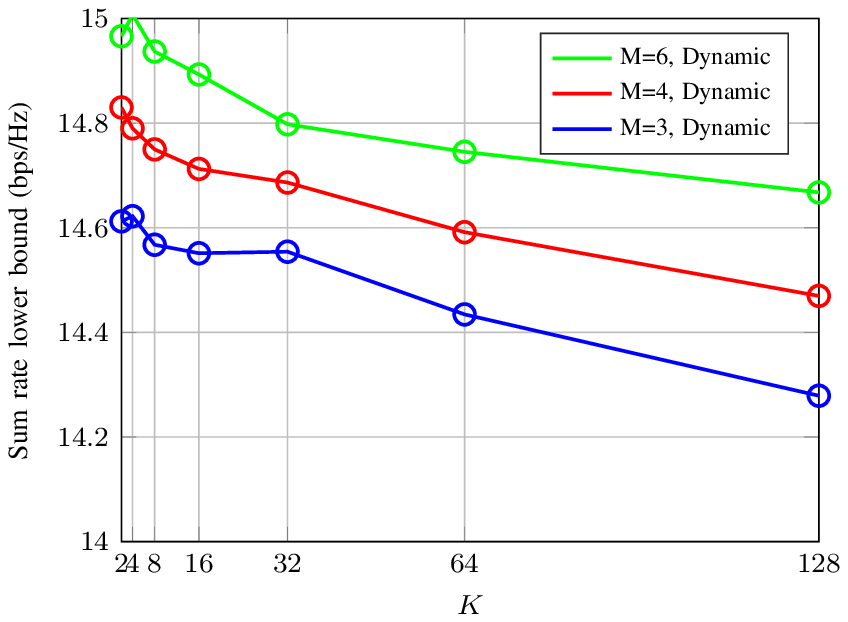}
    \end{minipage}%
    \begin{minipage}[t]{0.5\linewidth}
        \centering
        \includegraphics[width=\columnwidth]{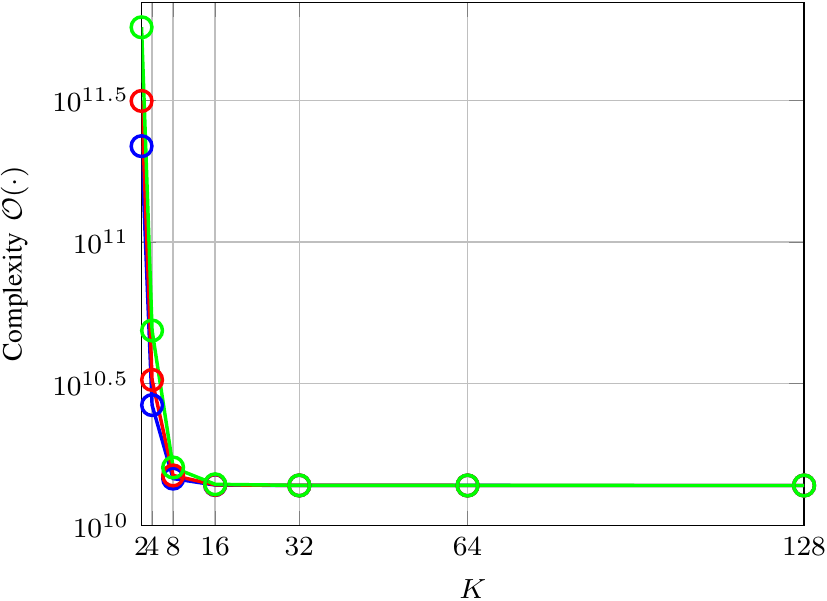}
    \end{minipage}\label{fig:choiceK}}
    \subfigure[$K=8$]{
        \begin{minipage}[t]{0.5\linewidth}
            \centering
            \includegraphics[width=\columnwidth]{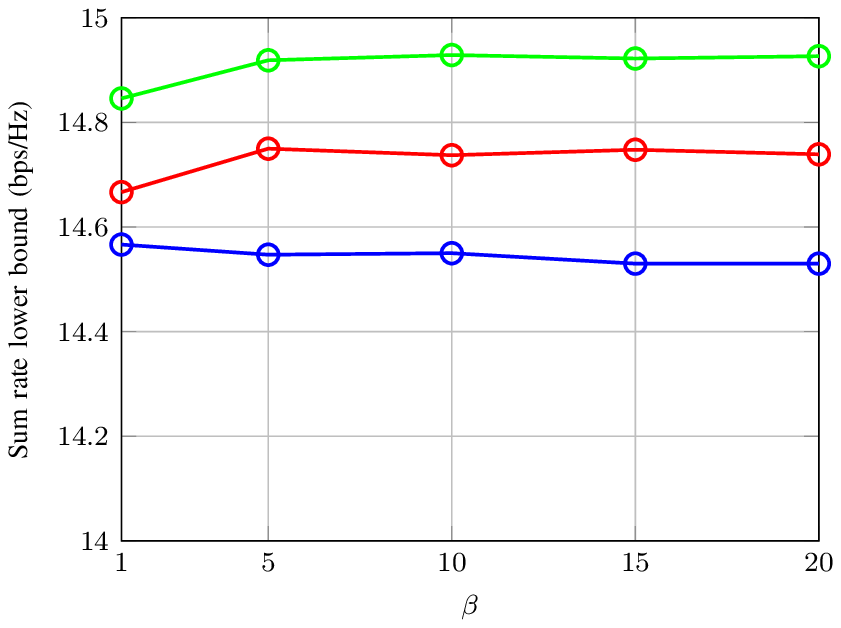}
        \end{minipage}%
        \begin{minipage}[t]{0.5\linewidth}
            \centering
            \includegraphics[width=\columnwidth]{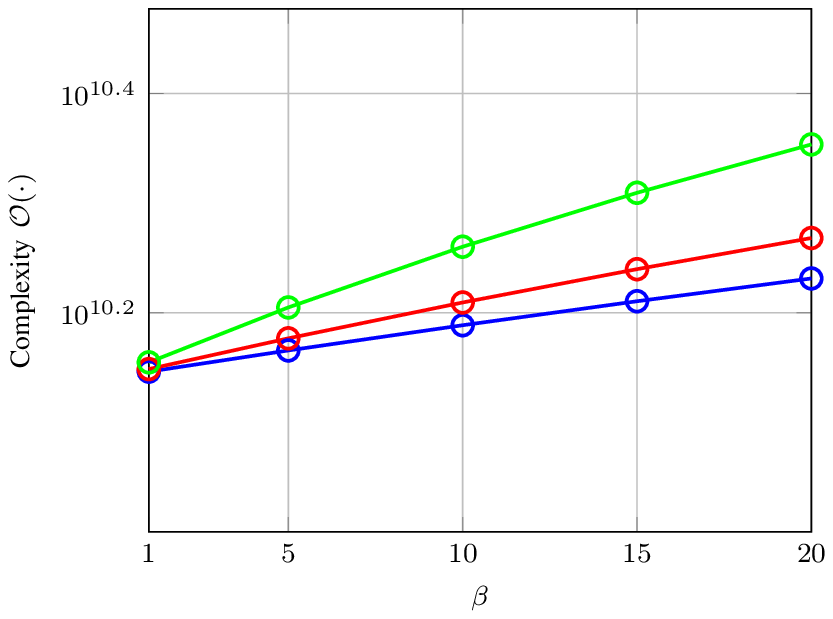}
        \end{minipage}\label{fig:choicebeta}}
    \label{fig:choice}
    \caption{The sum rate performance ($N=4$) and detection complexity in one transmission block (containing 100 symbols) as a function of $K$ and $\beta$ with $l=4, N_r=64, M'=2$.}
\end{figure}

The SL-FS algorithm has the large dimensional matrix
$\mathbf{\Delta}_B$ partitioned into $K$ low dimensional submatrices
$\mathbf{\Delta}_{B_k}$ for computational purposes. In this
subsection, we discuss further the impact of $K$ on the system
performance. Specifically, $K$ is chosen as the divisor of $M'NN_r$,
namely $\mod(M'NN_r,K)=0$. Fig. \ref{fig:choiceK} shows the sum rate
performance and the detection complexity by using the simplified
SL-RGS dynamic oversampling technique as a function of $K$, where
both decrease with the increase of $K$. These results indicate that
the parameter $K$ should be chosen properly in order to obtain
satisfactory sum rate performance and low detection complexity.
Moreover, Fig. \ref{fig:choicebeta} shows the sum rate and the
detection complexity as a function of $\beta$. In the SL-RGS
algorithm, the pre-defined parameter $\beta$ controls the search
range in each row. From the results, it is seen that beyond
$\beta=5$ there are diminishing returns on the sum rate
performance.

\subsection{Power Consumption}

\begin{figure}[!htbp]
    \centering
    \includegraphics{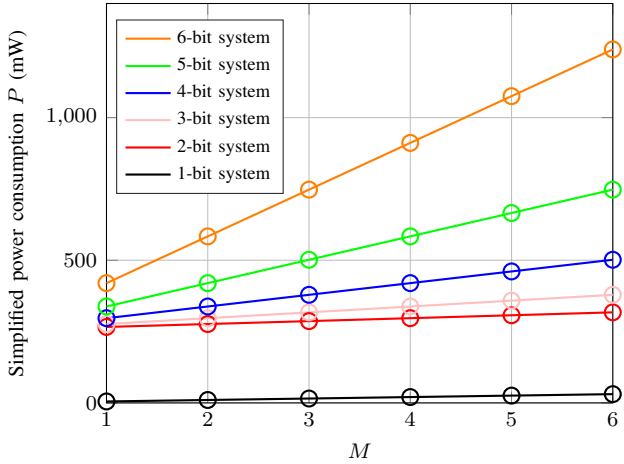}
    \caption{Simplified power consumption at the receiver with $N_r=64$.}
    \label{fig:Energy}
\end{figure}

Here, we investigate the receiver power consumption of systems with
uniform and dynamic oversampling versus the number of bits used by
the ADCs. Fig. \ref{fig:Energy} shows the simplified power
consumption $P$ at the receiver as a function of the quantization
bits $b$, where only the power consumption of different
components among different systems is considered. The simplified power consumption is calculated as
\begin{equation}
P = 2N_r(cP_{\text{AGC}}+P_{\text{ADC}}),
\end{equation}
where $P_{\text{AGC}}$ denotes the power consumption of AGC. $c$ is
chosen as 0 for 1-bit system and 1 for systems with more-bits. The
$P_{\text{ADC}}$ is given by
\begin{equation}
P_{\text{ADC}} = \text{FOM}_w\times Mf_{\text{Nyquist}}\times 2^{b},\quad b=1,2,\cdots
\end{equation}
where $f_{\text{Nyquist}}$ is the Nyquist-sampling rate. Numerical
parameters are based on \cite{3549}, where $P_{\text{AGC}}=2$mW,
$\text{FOM}_w$ is 200 fJ/conversion-step at 50 MHz bandwidth and
$f_{\text{Nyquist}}$ is 100 MHz. Fig.
\ref{fig:Energy} shows the simplified power consumption as a
function of the sampling rate, where the 1-bit system consumes the
least power.

\section{Numerical Results}

In this section, the uplink of a large-scale MIMO system
\cite{mmimo,wence} with $N_r=64$ and $N_t=4$ users is considered.
The $m(t)$ and $p(t)$ are normalized root-raised-cosine (RRC)
filters with a roll-off factor of 0.8 and the time delay $n^d_{n_t}$
of each terminal is uniformly distributed between $-T$ and $T$. The
channel is assumed to experience Rayleigh block fading. The
simulation results presented here are obtained by averaging over
1000 independent realizations of the channel matrix $\mathbf{H'}$,
noise and symbol vectors. The SNR is defined as
$10\log(\frac{N_t}{\sigma_n^2})$. In systems with dynamic
oversampling \cite{dynovs}, the performance of different initial
sampling rates are compared under the condition that the signal
processing rates $M'$ are equal to 2. In the SL-BFS and SL-RGS
algorithms $\beta=5$ and $K=8$.

\subsection{Design based on the sum rate criterion}

\begin{figure}[!htbp]
    \centering
    \subfigure[GEVD]{
    \begin{minipage}[t]{0.5\linewidth}
        \centering
        \includegraphics[width=\columnwidth]{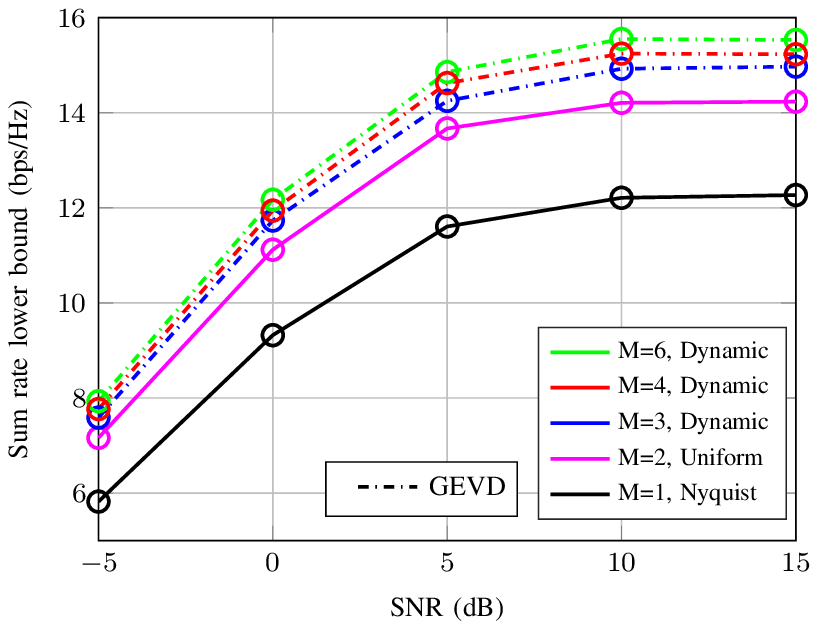}
        \label{fig:Capacity}
    \end{minipage}}\subfigure[SL-FS]{
    \begin{minipage}[t]{0.5\linewidth}
        \centering
        \includegraphics[width=\columnwidth]{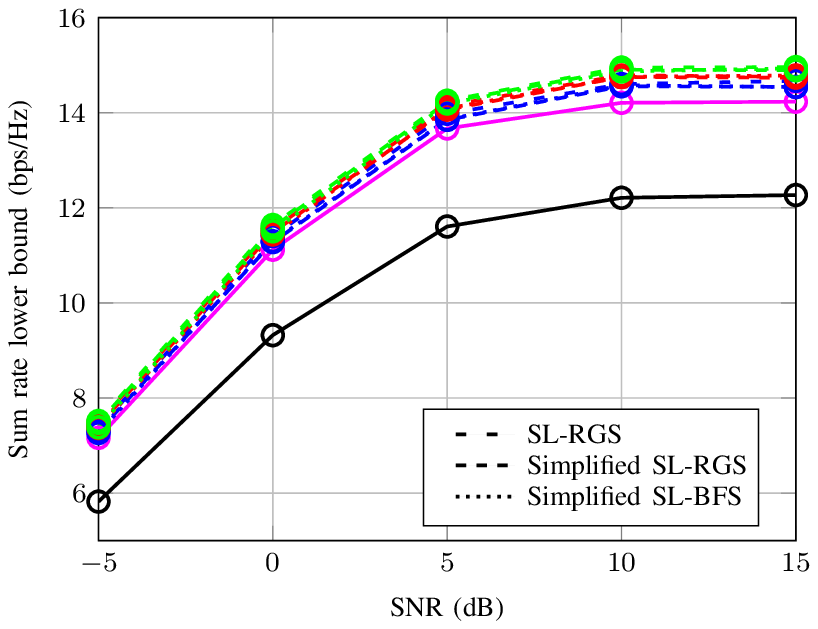}
        \label{fig:Capacity2}
    \end{minipage}}
    \caption{Comparison among Nyquist sampling, uniform oversampling and GEVD and SL-FS based dynamic oversampling with $M'=2$.}
\end{figure}

We first consider the system design based on the sum rate criterion,
where each transmission block contains 4 symbols and Gaussian
signaling is assumed. Fig. \ref{fig:Capacity} shows the sum rate
performance of systems with uniform and GEVD based dynamic
oversampling, where the latter outperforms the former.
Fig. \ref{fig:Capacity2} compares the performance of SL-RGS based dynamic
oversampling technique with standard and simplified version. As
depicted in Fig. \ref{fig:Capacity2}, the performance of the
simplified and the standard SL-RGS algorithms are comparable, and
dynamic oversampling with both versions outperforms uniform
oversampling. In terms of complexity, the proposed simplified SL-FS
algorithms have a much lower computational cost than the standard
SL-FS algorithms, while their performance are comparable. Fig.
\ref{fig:Capacity2} also compares the performance of the simplified SL-FS
algorithms, where both SL-BFS and SL-RGS achieve similar sum rates
\footnote{The performance of the standard SL-BFS algorithm is not
    shown in Fig. \ref{fig:Capacity2} due to the very high computational
    cost shown in subsection \ref{sec_complexity}.}. Figs.
\ref{fig:Capacity} and \ref{fig:Capacity2} show
that among the compared dimension reduction algorithms the GEVD
outperforms the SL-FS algorithms, since more information is
exploited and the samples are combined, but the simplified SL-FS
algorithms require lower computational costs. The green lines
(Dynamic, $M=6$, $M'=2$) in the figures indicate that a sum-rate
performance gain of up to 10\% over that of uniform oversampling
with $M=2$ can be achieved.

\subsection{Design based on the MSE criterion}

In this subsection, we examine the proposed dynamic oversampling
techniques in terms of the normalized MSE and SER. The modulation
scheme is quadrature phase-shift keying (QPSK). Each transmission
block contains 100 symbols. The window length $l$ is chosen as 4
Nyquist-sampled symbols.

\begin{figure}[!htbp]
    \centering
    \subfigure[GEVD]{
        \begin{minipage}[t]{0.5\linewidth}
            \centering
            \includegraphics[width=\columnwidth]{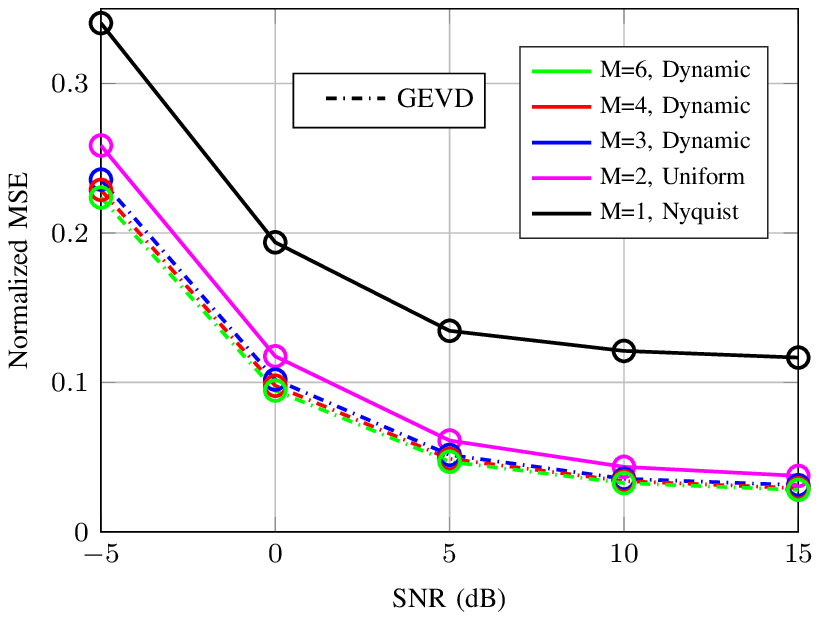}
            \label{fig:NMSE}
    \end{minipage}}\subfigure[SL-FS]{
        \begin{minipage}[t]{0.5\linewidth}
            \centering
            \includegraphics[width=\columnwidth]{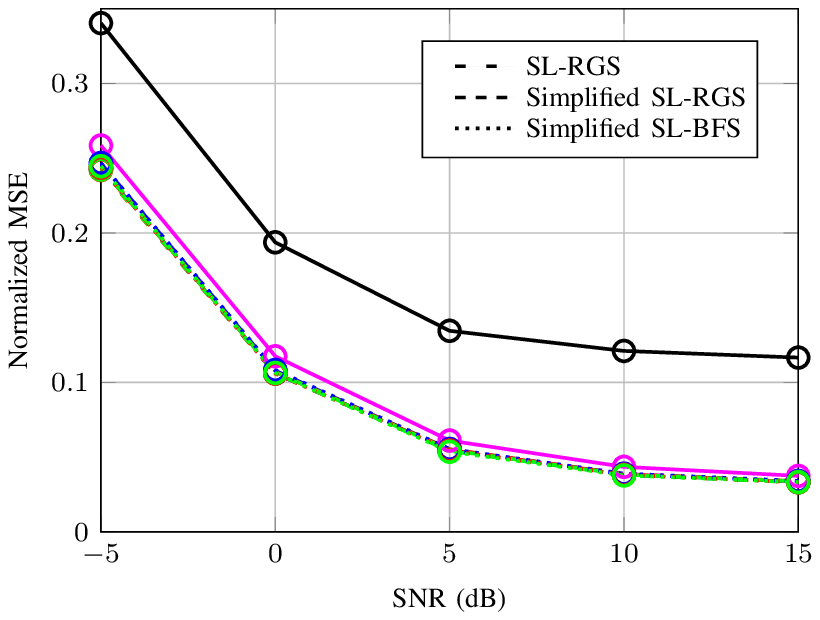}
            \label{fig:NMSE2}
    \end{minipage}}
    \caption{Comparison among Nyquist sampling, uniform oversampling and GEVD and SL-FS based dynamic oversampling with $M'=2$.}
\end{figure}

Figs. \ref{fig:NMSE} and \ref{fig:NMSE2} show the normalized MSE
performance as a function of SNR with GEVD based dynamic
oversampling, respectively. The results show that dynamic
oversampling outperforms uniform oversampling with $M = 2$ while
making the signal detection.
In terms of complexity, Fig. \ref{fig:complexity2} shows the
GEVD based dynamic oversampling has
higher complexity than uniform oversampling with $M=2$. This reveals
the complexity disadvantages of the GEVD based dynamic oversampling
technique.

\begin{figure}[!htbp]
    \centering
    \includegraphics{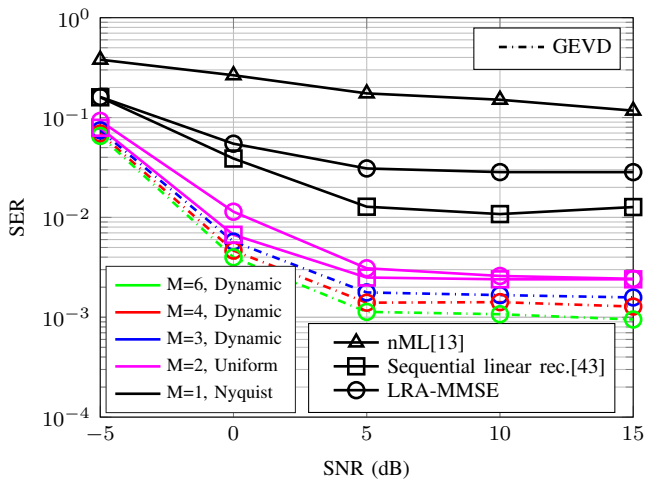}
    \caption{Comparison among Nyquist sampling, uniform oversampling and GEVD based dynamic oversampling with $M'=2$.}
    \label{fig:ser}
\end{figure}

\begin{figure}[!htbp]
    \centering
    \includegraphics{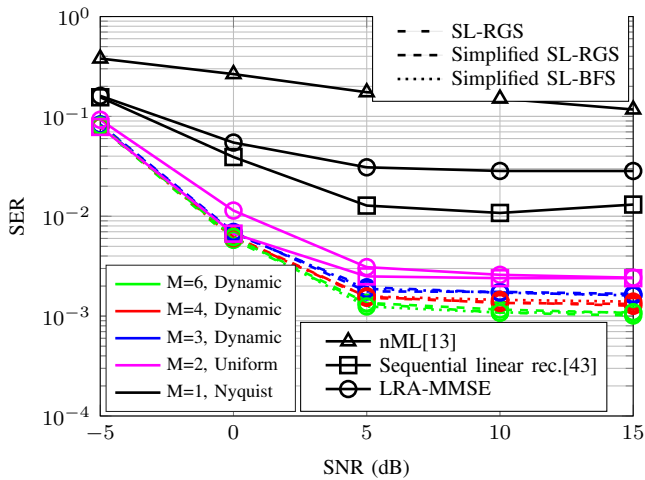}
    \caption{Comparison among Nyquist sampling, uniform oversampling and SL-FS based dynamic oversampling technique with $M'=2$.}
    \label{fig:ser2}
\end{figure}

Furthermore, Figs. \ref{fig:ser} and \ref{fig:ser2} show the SER
performance of the GEVD and SL-FS based dynamic oversampling
technique, respectively. It can be seen that the SL-RGS based
dynamic oversampling outperforms uniform oversampling with $M=2$
without increasing the computational cost. This reveals the
performance advantages of the proposed SL-RGS based dynamic
oversampling techniques. Moreover, in Fig. \ref{fig:ser2} the SER
performance of the simplified SL-BFS and the simplified SL-RGS
algorithms is also compared. The simplified SL-BFS and the
simplified SL-RGS with $M'=2$ have similar performance, whereas they
significantly outperform uniform oversampling. Figs. \ref{fig:ser}
and \ref{fig:ser2} have also compared the SER performance of the
proposed LRA-MMSE detector and dynamic oversampling scheme with
other existing 1-bit detectors with sampling at the Nyquist rate
\cite{7439790} and uniform oversampling \cite{8690724}, where the
former achieves the best performance.

\begin{figure}[!htbp]
    \centering
    \includegraphics{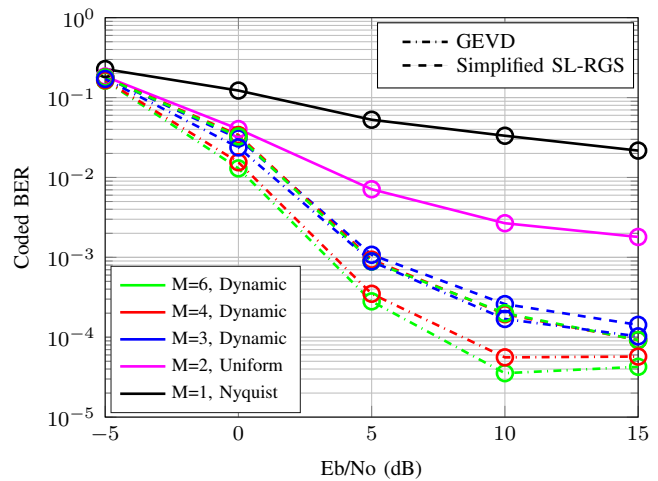}
    \caption{Comparison between GEVD and simplified SL-RGS based dynamic oversampling with $M'=2$.}
    \label{fig:ber}
\end{figure}

In order to further investigate the practical benefits of the
proposed dynamic oversampling technique, we have compared the
analyzed systems using $N_t=12$, $N_r=32$ and Low-Density
Parity-Check (LDPC) codes with block length $n = 512$ and rate
$R=1/2$ \cite{memd} using an iterative detection and decoding scheme
\cite{Shao2}. Other iterative detection and decoding schemes
\cite{spa,mfsic,mfdf,mbdf,did,bfidd,rrser,listmtc} are also
possible. Fig. \ref{fig:ber} shows the bit error rate (BER)
performance of the coded systems, where Eb/No is defined as
$10\log(\frac{N_tR}{\sigma_n^2})$. The results show that the
proposed dynamic oversampling outperforms the existing uniform
oversampling and Nyquist-rate techniques.

\section{Conclusions}

This work has proposed dynamic oversampling techniques for
large-scale multiple-antenna systems with 1-bit quantization at the
receiver. We have developed designs based on the sum rate and the
MSE criteria. Dimension reduction algorithms have also been devised
based the GEVD and sparse binary matrices in order to perform
dynamic oversampling. Furthermore, the proposed techniques have been analyzed
in terms of convergence, computational complexity and
power consumption at the receiver. Simulation results have shown
that the proposed dynamic oversampling can outperform uniform
oversampling in terms of sum rate, MSE or SER performance but with
much lower computational costs.

\bibliographystyle{IEEEtran}
\bibliography{ref}

\end{document}